\documentclass[twocolumn,showpacs,preprintnumbers,amsmath,amssymb]{revtex4}
\usepackage{graphicx}% Include figure files
\usepackage{dcolumn}% Align table columns on decimal point
\usepackage{bm}% bold math

\newcommand{\mean}[1]{\langle{#1}\rangle}

\newcommand{\ket}[1]{|{#1}\rangle}

\newcommand{\dgg}{^{\dagger}}
\newcommand{\trans}{^{{\mathsf T}}}
\newcommand{\Tr}{{\rm Tr}\hspace{0.07cm}}

\newcommand{\im}{{\rm i}}

\newcommand{\half}{\frac{1}{2}}

\begin{document}

\preprint{APS/123-QED}

\title{
Avoiding entanglement sudden-death via measurement feedback control
in a quantum network
}% Force line breaks with \\

%%%% Author %%%%
%%% Naoki Yamamoto, Hendra I. Nurdin, Matthew R. James, Ian R. Petersen %%%

\author{Naoki Yamamoto}
 \email{yamamoto@appi.keio.ac.jp}
\affiliation{%
Department of Applied Physics and Physico-Informatics,
Keio University, Yokohama 223-8522, Japan
}
\author{Hendra I. Nurdin}
 \email{hendra.nurdin@anu.edu.au}
\author{Matthew R. James}
 \email{matthew.james@anu.edu.au}
\affiliation{%
Department of Engineering, Australian National University,
Canberra, ACT 0200, Australia
}
\author{Ian R. Petersen}
 \email{i.petersen@adfa.edu.au}
\affiliation{%
School of Information Technology and Electrical Engineering,
University of New South Wales at the Australian Defence Force Academy,
Canberra, ACT 2600, Australia
}

\date{\today}% It is always \today, today,

\begin{abstract}

In this paper, we consider a linear quantum network composed of
two distantly separated cavities that are connected via a one-way
optical field.
When one of the cavity is damped and the other is undamped,
the overall cavity state obtains a large amount of entanglement
in its quadratures.
This entanglement however immediately decays and vanishes
in a finite time.
That is,  entanglement sudden-death occurs.
We show that the direct measurement
feedback method proposed by Wiseman can avoid this entanglement
sudden-death, and further, enhance the entanglement.
It is also shown that the entangled state under feedback control
is robust against signal loss in a realistic detector,
indicating the reliability of the proposed direct feedback method
in practical situations.

\end{abstract}

\maketitle

%%%%%%%%%%%%%%%%%%%%%%%%%%%%%%%%%%%%%%%%%%%%%%%%%%%%%%%%%%%%%%%%%%%%%%%%%
%%%%%%%%%%%%%%%%%%%%%%%%%%% Introduction %%%%%%%%%%%%%%%%%%%%%%%%%%%%%%%%
%%%%%%%%%%%%%%%%%%%%%%%%%%%%%%%%%%%%%%%%%%%%%%%%%%%%%%%%%%%%%%%%%%%%%%%%%

\section{Introduction}

Reliable generation and distribution of entanglement
in a quantum network is a central subject in quantum
information technology \cite{nielsen}, especially in
quantum communication \cite{Cirac,vanEnk,Parkins1,
Parkins2}.
The biggest issue in such systems is the decay of entanglement
due to decoherence effects that are inevitably
introduced when node-channel or channel-environment
interaction occurs.
Entanglement distillation \cite{Bennett1,Bennett2} is a useful
technique that restores such degraded entanglement. However, it
sometimes happens that entanglement completely disappears in a
finite time, which is called {\it entanglement sudden-death}
\cite{Eberly1,Eberly2}.
In this case, distillation techniques cannot recover
the vanished entanglement.

On the other hand,  feedback control can
be used to modify the dynamical structure of a system and improve
its performance, e.g., see \cite{Doherty1,Thomsen, Ahn,Bouten}.
Entanglement protection or generation is one of the most
attractive applications of feedback \cite{Wang,Carvalho,
Mirrahimi,Yamamoto}. In particular, two studies have demonstrated
that a feedback controller effectively assists in the distribution
of entanglement in a quantum network. One such result is by
Mancini and Wiseman \cite{Mancini}, where a direct measurement
feedback method \cite{Wiseman1, Wiseman2} is used to enhance the
correlation of two bosonic modes that couple through a
$\chi^{(2)}$ nonlinearity. The other such result is by Yanagisawa
\cite{Yanagisawa}, where an estimation-based feedback controller
is used to deterministically generate an entangled photon number
state of two distantly separated cavities.

This paper follows a similar direction to \cite{Mancini}
and \cite{Yanagisawa}.
That is, we also consider a problem of distributing
entanglement in a quantum network using direct feedback control.
The quantum network being considered is depicted in Fig.~1:
Two spatially  separated cavities are connected via a one-way
optical field, and the measurement results of the output of
Cavity~2 are directly fed back to control both cavities.
A more specific description will be given in Sections II-B
and II-C, but here we note that the network model considered
brings together the following three features that have been
not simultaneously considered in  previous work.
First, the network contains models of realistic components;
a realistic quantum channel in contact with an environment
and a realistic homodyne detector with finite bandwidth
\cite{Warszawski1,Warszawski2,Warszawski3}.
A realistic model is of practical importance for real-time
quantum feedback control.
Second, we consider linear continuous-variable cavity models
(i.e., we consider the quadratures of the cavity mode),
similar to the case of \cite{Parkins1,Parkins2,Mancini}.
Hence, the system differs from a discrete-variable system
such as an atomic energy level \cite{Cirac,vanEnk}, or
a photon number system description \cite{Yanagisawa}.
This setup is motivated by the recent rapid progress and deep
understanding of continuous-variable systems in the quantum
information regime \cite{Braunstein}.
Third, the cavities are spatially separated, and the
interaction between them is simply mediated by an optical field,
while in \cite{Mancini} two bosonic modes interact
through a $\chi^{(2)}$ optical nonlinearity and thus
the two modes are not spatially separated.
The spatially separated case is the case of  interest
in applications such as quantum communication.

\begin{figure}[t]
\centering

\begin{picture}(220,140)
\put(1,7)
{\includegraphics[width=3in]{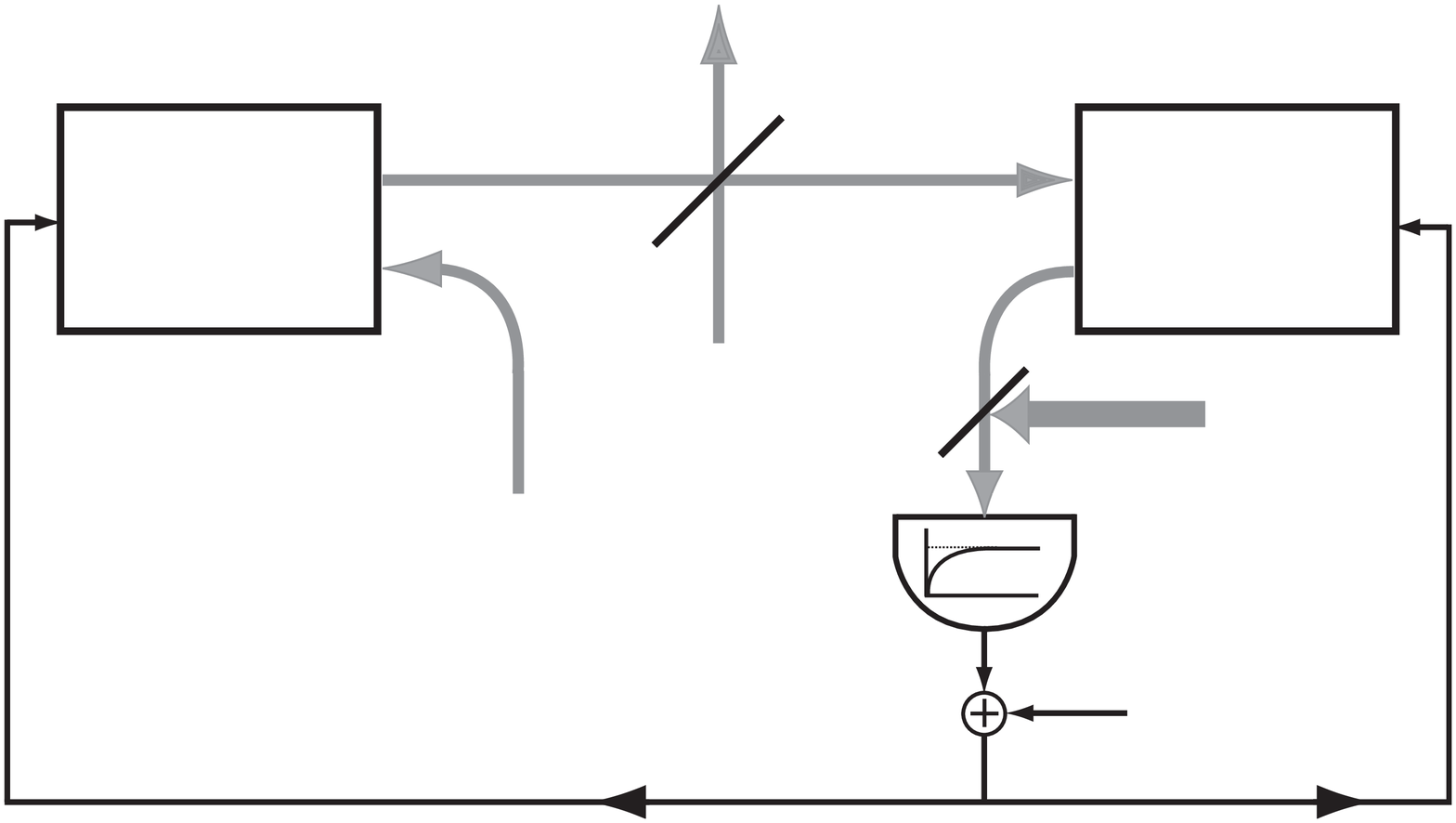}}
%\put(1,7)
%{\includegraphics[width=3in]{CavityCascade.png}}
\put(18,118){Cavity 1}
\put(169,118){Cavity 2}
\put(31,93){$\hat{a}_1$}
\put(182.5,93){$\hat{a}_2$}
\put(69,44){$\hat{B}_{1,t}$}
\put(113,76){$\hat{B}_{2,t}$}
\put(165,27){$\hat{B}_{3,t}+\hat{B}_{3,t}^*$}
\put(185,65){LO}
\put(112,40){LPF}
\put(120,116){BS}
\put(136,14){$y_t$}
\put(92,1){$g$}
\put(198,1){$g$}
\end{picture}

\caption{ Schematic of the network.
Thick gray lines represent
quantum optical fields, while thin black lines represent
classical signals. } \label{fg_cav_cas}
\end{figure}

The contributions of this paper are as follows.
First, we show that the network considered in this paper,
which looks complicated, can be systematically captured
and described using the theory of quantum cascade systems
\cite{Carmichael1,Carmichael2,Gardiner,Gough}.
We then show that when the first cavity is undamped and
the second cavity is damped, the
cavity state obtains a large amount of entanglement, which,
however, disappears in a finite time despite the continuous
interaction between the cavities; i.e., entanglement
sudden-death occurs.
As mentioned above, no distillation technique can recover
such a zero entanglement.
Nevertheless, we show that direct feedback control
not only prevents entanglement sudden-death, but can also
enhance the entanglement.
Moreover, it will be shown that the entangled state under
control is robust against signal loss in a realistic detector,
implying the reliability of the direct feedback
method in practical situations.

We use the following notation.
For a matrix $A=(a_{ij})$, the symbols $A^{{\mathsf T}}$,
$A\dgg$, and $A^*$ represent its transpose, conjugate transpose,
and elementwise complex conjugate of $A$, i.e.,
$A^{{\mathsf T}}=(a_{ji})$, $A\dgg=(a_{ji}^*)$,
and $A^*=(a_{ij}^*)=(A\dgg)^{{\mathsf T}}$, respectively.
These rules are applied to any rectangular matrix including
column and row vectors.
${\rm Re}(A)$ and ${\rm Im}(A)$ denote the real and imaginary
part of $A$, respectively, i.e., $({\rm Re}(A))_{ij}=(a_{ij}+a_{ij}^*)/2$
and $({\rm Im}(A))_{ij}=(a_{ij}-a_{ij}^*)/2\im$.
The matrix element $a_{ij}$ can be an operator $\hat{a}_{ij}$;
in this case, $\hat{a}_{ij}^*$ denotes its adjoint operator.

%%%%%%%%%%%%%%%%%%%%%%%%%%%%%%%%%%%%%%%%%%%%%%%%%%%%%%%%%%%%%%%%%%%%%%%%%
%%%%%%%%%%%%%%%%%%%%%%%%%%%%%% General model %%%%%%%%%%%%%%%%%%%%%%%%%%%%
%%%%%%%%%%%%%%%%%%%%%%%%%%%%%%%%%%%%%%%%%%%%%%%%%%%%%%%%%%%%%%%%%%%%%%%%%

\section{Model}

\subsection{General linear quantum systems}
We consider  a general linear continuous-variable system with
$N$-degrees of freedoms. Let $\hat{{\bf x}}_i=(\hat{q}_i,
\hat{p}_i)^{{\mathsf T}}$ be the standard quadrature pair of the
$i$-th subsystem. It then follows from the canonical commutation
relation
$[\hat{q}_i, \hat{p}_j]
=\hat{q}_i\hat{p}_j-\hat{p}_j\hat{q}_i =\im\delta_{ij}~(\hbar=1)$
that the vector of system variables $\hat{{\bf x}} :=(\hat{{\bf
x}}_1\trans,\ldots,\hat{{\bf x}}_N\trans)\trans$ satisfies
\[
   \hat{{\bf x}}\hat{{\bf x}}\trans
      -(\hat{{\bf x}}\hat{{\bf x}}\trans)\trans
         =\im\Sigma_N
         =\im\bigoplus_{k=1}^N\Sigma,~~~
   \Sigma:=\left[ \begin{array}{cc}
            0 & 1 \\
            -1 & 0 \\
           \end{array} \right].
\]
Suppose that the system contacts with $M$ optical vacuum
fields without scattering.
In general, such an interaction is described by a unitary
operator
that obeys the {\it Hudson-Parthasarathy equation} \cite{Hudson}:
\begin{eqnarray}
& & \hspace*{-1em}
\label{HP-eq}
   d\hat{U}_t
      =\Big[ \big(-\im \hat{H}
                  -\half\sum_{k=1}^{M}\hat{L}_k^* \hat{L}_k\big)dt
\nonumber \\ & & \hspace*{1.5em}
       \mbox{}
       +\sum_{k=1}^{M}\big(\hat{L}_kd\hat{B}_{k,t}^*
                  -\hat{L}_k^* d\hat{B}_{k,t}\big) \Big]\hat{U}_t,~~~
   \hat{U}_0=\hat{I}.
\end{eqnarray}
The operators $\hat{B}_{k,t}$ and $\hat{B}_{k,t}^*$ represent
the quantum annihilation and creation processes on the $k$-th
field, respectively.
Note that $[d\hat{B}_{i,t},d\hat{B}^*_{j,t}]=\delta_{ij}dt$.
Let us choose $\hat{H}=\hat{H}^*$ and $\hat{L}_k$
as follows:
\begin{equation}
\label{general Hamiltonian}
    \hat{H}=\half\hat{{\bf x}}\trans G\hat{{\bf x}},~~~
    \hat{L}_k=L_k\trans\hat{{\bf x}},
\end{equation}
where $G=G\trans\in{\mathbb R}^{2N\times 2N}$ and
$L_k\in{\mathbb C}^{2N}$.
The system variables obey the Heisenberg equation
$\hat{{\bf x}}_t:=(\ldots,\hat{U}_t^*\hat{q}_i\hat{U}_t,
\hat{U}_t^*\hat{p}_i\hat{U}_t,\ldots)\trans$.
We then obtain the following linear equation:
\begin{equation}
\label{linear-qsde}
    d\hat{{\bf x}}_t=A\hat{{\bf x}}_tdt
           +\im\Sigma_N[L^{{\mathsf T}}d\hat{{\bf B}}_t^*
                     -L\dgg d\hat{{\bf B}}_t],
\end{equation}
where $L\trans:=(L_1,\ldots,L_M)\in{\mathbb C}^{2N\times M}$,
$A:=\Sigma_N[G+{\rm Im}(L\dgg L)]$, and 
$\hat{{\bf B}}_t:=(\hat{B}_{1,t},\ldots,\hat{B}_{M,t})\trans$. 
For details on the physical meaning of these abstract linear models, 
see Section III and, e.g., \cite{Edwards,JNP06,NJD08}. 
It is easy to see that the first moment vector 
$\mean{\hat{{\bf x}}_t}
:=(\ldots,\mean{\hat{U}_t^*\hat{q}_i\hat{U}_t},
\mean{\hat{U}_t^*\hat{p}_i\hat{U}_t},\ldots)\trans$, where
$\mean{\hat{x}}:=\Tr(\hat{\rho}\hat{x})$, satisfies the linear
equation $d\mean{\hat{{\bf x}}_t}/dt=A\mean{\hat{{\bf x}}_t}$.
Also, the covariance matrix $V_t=(\mean{\hat{{\bf V}}_{ij}})$,
where
\[
   \hat{{\bf V}}
         =\half\Big[
             \Delta\hat{{\bf x}}_t\Delta\hat{{\bf x}}_t^{{\mathsf T}}
            +(\Delta\hat{{\bf x}}_t\Delta\hat{{\bf x}}_t^{{\mathsf T}})
                                                            ^{{\mathsf T}}
                           \Big],~~~
   \Delta\hat{{\bf x}}_t:=\hat{{\bf x}}_t-\mean{\hat{{\bf x}}_t},
\]
satisfies the following {\it Lyapunov matrix differential equation}:
\begin{equation}
\label{Lyapunov Eq}
   \frac{dV_t}{dt}=AV_t+V_tA\trans+D.
\end{equation}
Here, $D:=\Sigma_N{\rm Re}(L\dgg L)\Sigma_N\trans$.
Suppose that the quantum state is Gaussian at $t=0$.
Then, from the linearity of the dynamics, the unconditional
state is always Gaussian with mean $\mean{\hat{{\bf x}}_t}$
and covariance $V_t$.
Note that the unconditional state corresponds to a classical
probability density that describes a linear diffusion
process.

%%%%%%%%%%%%%%%%%%%%%%%%%%%%%%%%%%%%%%%%%%%%%%%%%%%%%%%%%%%%%%%%%%%%%%%%%
%%%%%%%%%%%%%%%%%%%%%%%%%% The ideal network %%%%%%%%%%%%%%%%%%%%%%%%%%%%
%%%%%%%%%%%%%%%%%%%%%%%%%%%%%%%%%%%%%%%%%%%%%%%%%%%%%%%%%%%%%%%%%%%%%%%%%

\subsection{The ideal network}

Before describing the quantum network depicted in Fig.~1, let us
consider the ideal situation where the optical field between the
cavities is not in contact with any environment, and the homodyne
detector is perfect. In this case, the system is the simple
cascade of two cavities with a feedback loop. The entangled state
of this ideal network will be compared to that of the realistic
one for the purpose of clarifying how much the realistic
parameters affect the system. Also, this ideal setup allows us to
determine a reasonable control Hamiltonian (the vector $f$ given
below), as will be seen in Section III-B.

Each component of the network is described as follows.
The optical vacuum field $\hat{B}_{1,t}$ comes into Cavity~1,
and then, its output becomes the input of Cavity 2.
We assume that, after some approximations, the $i$-th
cavity-field interaction is represented by Eq. \eqref{HP-eq}
with single field $(M=1)$ and with the following operators:
\[
    \hat{H}_i=\half\hat{{\bf x}}_i^{{\mathsf T}}G_i\hat{{\bf x}}_i,~~~
    \hat{L}_{1,i}=\ell_i^{{\mathsf T}}\hat{{\bf x}}_i,~~~
    (i=1,2),
\]
where $\hat{q}_i=(\hat{a}_i+\hat{a}_i^*)/\sqrt{2}$ and
$\hat{p}_i=(\hat{a}_i-\hat{a}_i^*)/\sqrt{2}\im$.
The subscript $(1,i)$ in $\hat{L}$ means the $1$-st field
and the $i$-th cavity (see the figures in Appendix~A).
Also, $G_i=G_i^{{\mathsf T}}\in{\mathbb R}^{2\times 2}$ and
$\ell_i\in{\mathbb C}^2$.
The output of Cavity 2 is transformed to a classical signal
$y_t$ through an ideal homodyne detector.
Suppose now that each cavity has an additional Hamiltonian
of the form
\begin{equation*}
     \hat{H}_i^{\rm fb}
       =u_t\hat{F}_i
          =u_tf_i\trans\hat{{\bf x}}_i,~~~
              f_i\in{\mathbb R}^2,~~~(i=1,2),
\end{equation*}
where $u_t\in{\mathbb R}$ is the control input.
Then, direct measurement feedback $u_t=gy_t$ closes the
loop by connecting the detector to the cavities.
Here $g\in{\mathbb R}$ is the control gain.
Note that we need a classical communication channel in order
to control Cavity~1;
that is, a local operation via classical communication
(LOCC) type control is performed.

For this network, we can easily determine the system matrices $G$
and $L_k$ in Eq. \eqref{general Hamiltonian} that specify the
whole dynamical equation. The derivation is based on the theory of
quantum cascade systems
\cite{Carmichael1,Carmichael2,Gardiner,Gough} and is given in
Appendix~A. Then, from the definition, the $A$ and $D$ matrices in
the Lyapunov equation \eqref{Lyapunov Eq} are readily obtained as
follows:
\begin{eqnarray}
& & \hspace*{-2em}
\label{ideal controlled A}
    A_{\rm id}=A_o+2g\Sigma_2 f{\rm Re}(\ell)\trans,
\\ & & \hspace*{-2.1em}
\label{ideal controlled D}
    D_{\rm id}=D_o
      +\Sigma_2\big[g^2ff\trans
           -g{\rm Im}(\ell)f\trans
           -gf{\rm Im}(\ell)\trans\big]\Sigma_2\trans,
\end{eqnarray}
where $\ell=(\ell_1\trans, \ell_2\trans)\trans,~
f=(f_1\trans, f_2\trans)\trans$, and
\begin{equation}
\label{ideal uncontrolled A and D}
    A_o
      =\left[ \begin{array}{cc}
          A_1 & 0 \\
          2\Sigma{\rm Im}(\ell_2^*\ell_1^{{\mathsf T}}) & A_2 \\
       \end{array} \right],~~
    D_o
      =\left[ \begin{array}{cc}
        D_1 & \star \\
        \Sigma{\rm Re}(\ell_2^*\ell_1\trans)\Sigma\trans & D_2 \\
       \end{array} \right].
\end{equation}
Here, $A_i=\Sigma[G_i+{\rm Im}(\ell_i^*\ell_i^{{\mathsf T}})]$,
$D_i=\Sigma{\rm Re}(\ell_i^*\ell_i^{{\mathsf T}})\Sigma\trans$,
and $\star$ denotes the symmetric elements.
Note that $A_o$ and $D_o$ are the system matrices of the network
without feedback.
Hence, the upper off-diagonal block matrix of $A_o$ is zero,
implying the one-way interaction of the cavities.

%%%%%%%%%%%%%%%%%%%%%%%%%%%%%%%%%%%%%%%%%%%%%%%%%%%%%%%%%%%%%%%%%%%%%%%%%
%%%%%%%%%%%%%%%%%%%%%%%% The realistic network %%%%%%%%%%%%%%%%%%%%%%%%%%
%%%%%%%%%%%%%%%%%%%%%%%%%%%%%%%%%%%%%%%%%%%%%%%%%%%%%%%%%%%%%%%%%%%%%%%%%

\subsection{The realistic network}

We are now in the position to describe a realistic network,
which introduces the following two assumptions.
First, the output of Cavity~1 is mixed with another vacuum
field $\hat{B}_{2,t}$ through a beam splitter (BS) with
transmittance $\alpha$.
This is a standard model of possible environmental effects
on a long quantum channel.
Second, the homodyne detector is not perfect and is described
by the one-dimensional classical dynamics
\begin{equation}
\label{detector}
   d\xi_t=a_1\xi_t dt+a_2dw_t,~~
     dy_t=a_3\xi_t dt+dv_t,~~
       a_i\in{\mathbb R},
\end{equation}
where $w_t$ is an input stochastic process satisfying
${\bf E}[dw_t^2]=dt$ and $v_t$ is an additional measurement
noise satisfying ${\bf E}[dv_t^2]=a_4dt~(a_4>0)$.
In particular, a typical low-pass filter (LPF) is realized by
choosing $a_i$ as
\[
    a_1=-\frac{1}{\tau},~~~a_2=\frac{1}{\tau},~~~a_3=1,
\]
where $\tau>0$ is the time-constant.
We now note that the detector \eqref{detector} can be
represented as a quantum system with two fields
$w_t=\hat{B}_{1,t}^{'}+\hat{B}_{1,t}^{'*}$ and
$v_t=\hat{B}_{3,t}+\hat{B}_{3,t}^*$, where $\hat{B}_{1,t}^{'}$
is the output of Cavity~2.
Indeed, from \cite{Gough}, Eq. \eqref{HP-eq} with $M=2$ and
with the operators
\[
    \hat{H}_3=\frac{a_1}{2}(\hat{q}_3\hat{p}_3+\hat{p}_3\hat{q}_3),~~~
    \hat{L}_{1,3}=-\im a_2\hat{p}_3,~~~
    \hat{L}_{3,3}=\frac{a_3}{2a_4}\hat{q}_3
\]
leads to a linear equation of the form \eqref{detector},
where $\hat{U}_t^*\hat{q}_3\hat{U}_t$ plays the same role of
$\xi_t$.

Consequently, the network is composed of two cavities, a beam
splitter, a detector, and a controller, with three optical vacuum
fields. (Note that the beam splitter with local oscillator
(LO) shown in Fig.~1 is a part of the detector.)
To systematically obtain the overall system matrices $G$ and
$L_k$ in Eq.~\eqref{general Hamiltonian} of this complicated
network, we again use the theory of quantum cascade systems
\cite{Carmichael1,Carmichael2,Gardiner,Gough}.
The procedure is given in Appendix~A.
We then obtain the matrices $A$ and $D$ in Eq.
\eqref{Lyapunov Eq} as follows:
\begin{eqnarray}
& & \hspace*{-1em}
\label{A matrix}
    A_{\rm re}
     =\left[ \begin{array}{ccc}
        A_1 & 0 & ga_3\Sigma f_1 \\
        2\alpha\Sigma{\rm Im}(\ell_2^*\ell_1^{{\mathsf T}}) & A_2
                                                 & ga_3\Sigma f_2  \\
        2\alpha a_2 {\rm Re}(\ell_1)\trans & 2 a_2{\rm Re}(\ell_2)\trans
                                                                  & a_1 \\
      \end{array} \right],
\\ & & \hspace*{-1em}
\label{D matrix}
    D_{\rm re}
     =\left[ \begin{array}{ccc}
        D_1 & \star & \star \\
        \alpha\Sigma{\rm Re}(\ell_2^*\ell_1\trans)\Sigma\trans
                                                      & D_2 & \star \\
        -\alpha a_2 {\rm Im}(\ell_1)\trans\Sigma\trans
                   & -a_2{\rm Im}(\ell_2)\trans\Sigma\trans & a_2^2 \\
      \end{array} \right]
\nonumber \\ & & \hspace*{2em}
     \mbox{}
     +g^2 a_4\left[ \begin{array}{ccc}
               \Sigma f_1f_1\trans\Sigma\trans & \star & 0 \\
               \Sigma f_2f_1\trans\Sigma\trans &
                     \Sigma f_2f_2\trans\Sigma\trans & 0 \\
               0 & 0 & 0 \\
            \end{array} \right].
\end{eqnarray}
%

%%%%%%%%%%%%%%%%%%%%%%%%%%%%%%%%%%%%%%%%%%%%%%%%%%%%%%%%%%%%%%%%%%%%%%%%%
%%%%%%%%%%%%%%%%%%%%%%%%% Entanglement Control %%%%%%%%%%%%%%%%%%%%%%%%%%
%%%%%%%%%%%%%%%%%%%%%%%%%%%%%%%%%%%%%%%%%%%%%%%%%%%%%%%%%%%%%%%%%%%%%%%%%

\section{Entanglement control}
\label{sec:entg-cntl}

In this section, we study the entanglement of the cavity state for
the ideal network. Since the state is Gaussian, its entanglement
is completely characterized by the covariance matrix
\cite{Duan,Simon}. In our case, the covariance matrix to be
evaluated is obtained by solving the Lyapunov equation
\eqref{Lyapunov Eq} with the coefficient matrices $A_{\rm id}$ and
$D_{\rm id}$ in Eqs. \eqref{ideal controlled A} and \eqref{ideal
controlled D}. Let the matrices $V_i$ be defined by the $2\times
2$ block matrix decomposition of $V$ as follows:
\begin{equation*}
\label{V deconposition}
    V=\left[ \begin{array}{cc}
            V_1 & V_2 \\
            V_2\trans & V_3 \\
      \end{array} \right].
\end{equation*}
Then, the following {\it logarithmic negativity} \cite{Vidal}
can be used as a reasonable measure of Gaussian entanglement:
\begin{equation}
\label{log negativity}
    E_{\cal N}={\rm max}\big\{0,~-\log(2\nu)\big\},
\end{equation}
where $\log x$ denotes the natural logarithm of $x$,
\begin{eqnarray}
& & \hspace*{-1em}
\label{symplectic eigenvalue}
    \nu:=\frac{1}{\sqrt{2}}
         \sqrt{ \tilde{\Delta}
              -\sqrt{ \tilde{\Delta}^2-4{\rm det}(V)} },
\\ & & \hspace*{-1em}
    \tilde{\Delta}:={\rm det}(V_1)+{\rm det}(V_3)-2{\rm det}(V_2).
\nonumber
\end{eqnarray}
The logarithmic negativity $E_{\cal N}$ divides the
state space into two regions: (i) the separable region,
corresponding to $E_{\cal N}=0$, and (ii) the entangled
region, within which $E_{\cal N} > 0$.
Thus phenomena of entanglement creation and destruction
can be understood simply in terms of movement of the system
between these two regions.

\subsection{Entanglement sudden-death}

Here we study the uncontrolled network; i.e., $g=0$.

We compute $E_{\cal N}$ for two situations.
First, consider the case where both cavities have the same
quadratic Hamiltonian and are damped as a result of the field-cavity
interaction, that is,
\[
    G_1=G_2=
      \left[ \begin{array}{cc}
        m & 0 \\
        0 & 1 \\
      \end{array} \right],~~~
    \ell_1=\ell_2=
      \sqrt{\kappa}
      \left[ \begin{array}{c}
        1 \\
        \im \\
      \end{array} \right],
\]
where $m>0$ and $\kappa>0$.
This type of quadratic Hamiltonian can be implemented in a
cavity system following the scheme of the {\it degenerate parametric
amplification} \cite{Gardiner}; see Appendix~B.
In this case, $A_{{\rm id}}$ is a {\it stable matrix}, and the
Lyapunov equation \eqref{Lyapunov Eq} has a unique steady state
solution (see e.g., \cite{Zhou}).
Now assume that at $t=0$, the cavity is in the separable state
satisfying $V_0=2I$.
When the optical field is switched on, the cavity modes
couple after a finite time (i.e., the ``entanglement sudden-birth"
\cite{Ficek} occurs), and a steady entangled state is generated
as seen from the dotted line in Fig.~2 (a).
However, in this case, the entanglement is very small
($E_{\cal N}\approx 0.21$).

This result can be understood by examining the trajectory of the
parameter $(\tilde{\Delta},{\rm det}(V))$. In Fig.~3, the colored
region with  contour lines represents the set of parameters
where a general two-mode Gaussian state is entangled, i.e.,
$E_{\cal N}>0$, while the white region corresponds to separable
states; i.e., $E_{\cal N}=0$.
The trajectory, denoted by ${\cal T}_{\rm dam}$, evolves toward
the steady entangled state that is located far from the area with
large $E_{\cal N}$ (the right bottom area in Fig.~3).
This is likely because each cavity has a strong tendency to
transit into the vacuum state due to the damping.
Indeed, when the cavity is in the separable vacuum state
$\ket{0}\ket{0}$, the corresponding covariance matrix satisfies
$(\tilde{\Delta}, {\rm det}(V))=(0.5, 0.0625)$, which is very
close to the equilibrium point of ${\cal T}_{\rm dam}$.
Moreover, Fig. 2 (b) shows that the purity (for a discussion
of physical meanings of the purity, see e.g. \cite{Paris}) of
the steady Gaussian state $\hat{\rho}$,
\begin{equation}
\label{purity}
    P:=\Tr(\hat{\rho}^2)
      =\frac{1}{4\sqrt{{\rm det}(V)}}\in(0,1],
\end{equation}
approaches $P \approx 0.8$ as $t\rightarrow\infty$.
This also suggests that the steady state is close to
the separable vacuum state.

The above observation motivates us to try a dispersive
field-cavity interaction, which results in a  phase shift of the output
field \cite{Doherty1,Milburn,Doherty2,Wiseman4}.
For a practical method to implement this kind of coupling in a
cavity system, see Appendix~B.
In this case, the cavity is not damped, and thus, it does not
have a tendency to move toward  the vacuum state.
In particular, we assume that only the first cavity has such an
interaction; i.e.,
\[
    \ell_1=(\sqrt{\kappa},~0)\trans,~~~
    \ell_2=\sqrt{\kappa}(1,~\im)\trans.
\]
We then find that $A_{\rm id}$ is not a stable matrix, and the
Lyapunov equation \eqref{Lyapunov Eq} need not have a steady state
solution as $t \rightarrow \infty$
\footnote{
More precisely, $A_{{\rm id}}$ is a {\it marginally stable
matrix}: the eigenvalues of $A_{{\rm id}}$, $\lambda_i~(i=1,2,3,4)$,
satisfy ${\rm Re}(\lambda_1)={\rm Re}(\lambda_2)=0$ and
${\rm Re}(\lambda_3), {\rm Re}(\lambda_4)<0$.
For a marginally stable matrix, the corresponding Lyapunov
equation need not have a steady solution.
%To see this in our case, let $\mu$ be the eigenvector of
%$A_{{\rm id}}\trans$ that corresponds to $\lambda_1$ or
%$\lambda_2$.
%We then readily find $d(\mu\dgg V_t\mu)/dt=\mu\dgg D\mu>0$,
%leading that $\mu\dgg V_t\mu\rightarrow\infty$ as $t\rightarrow\infty$.
}.
 Fig.~3 shows that the corresponding
trajectory, denoted by ${\cal T}_{\rm dis}$, evolves far from
the separable initial state and reaches the area with large
$E_{\cal N}$.
This figure also shows how both the entanglement and
purity decreases as time goes on and ${\cal T}_{\rm dis}$
escapes from the region of entangled states at $t=6.2$.

Finally, we remark that, if we exchange the order of the
interactions, i.e., $\ell_1=\sqrt{\kappa}(1,~\im)\trans$
and $\ell_2=(\sqrt{\kappa},~0)\trans$, the corresponding
trajectory remains within the region of separable states,
i.e., $E_{\cal N}=0$ for all $t>0$.
The situation is much the same even when each cavity interacts
with the field in a dispersive way.

\begin{figure}[t]
\centering

\begin{picture}(170,145)
\put(-1,1)
{\includegraphics[width=2.4in]{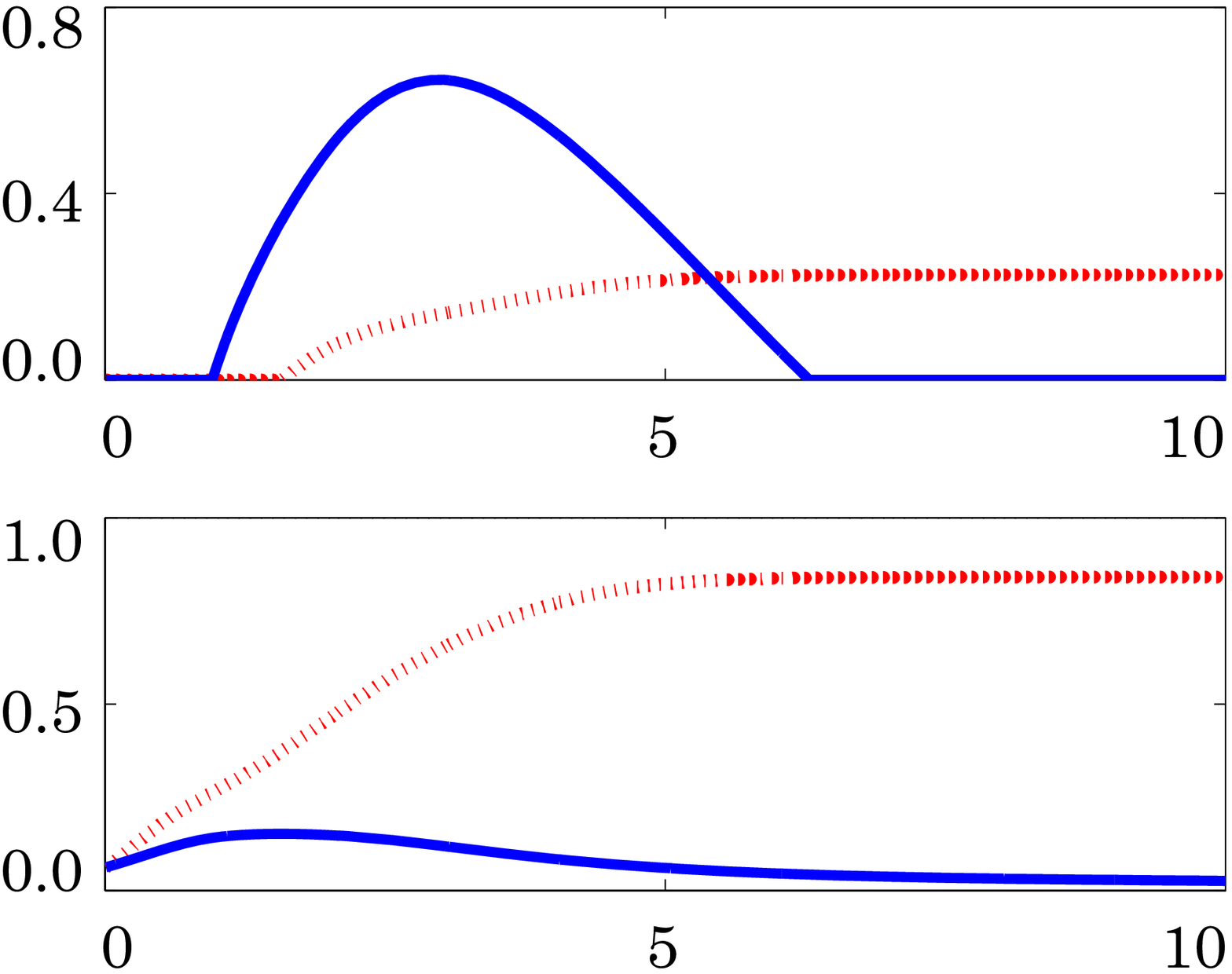}}
%\put(-3,1)
%{\includegraphics[width=2.6in]{Sudden2.png}}
\put(52,1){$t$}
\put(52,71){$t$}
\put(106,73){{\footnotesize 6.2}}
\put(-15,115){$E_{\cal N}$}
\put(-12,45){$P$}
\put(172,123){(a)}
\put(172,54){(b)}
\end{picture}

\caption{ Time-dependence of the logarithmic negativity
(a) and the purity (b) of the overall cavity state without
feedback control.
The solid and dotted lines correspond to the dispersive-damped
and damped-damped cases, respectively.
The parameters are $m=0.2$ and $\kappa=1$. }
\end{figure}

\begin{figure}[t]
\centering

\begin{picture}(170,155)
\put(-6,1)
{\includegraphics[width=2.6in]{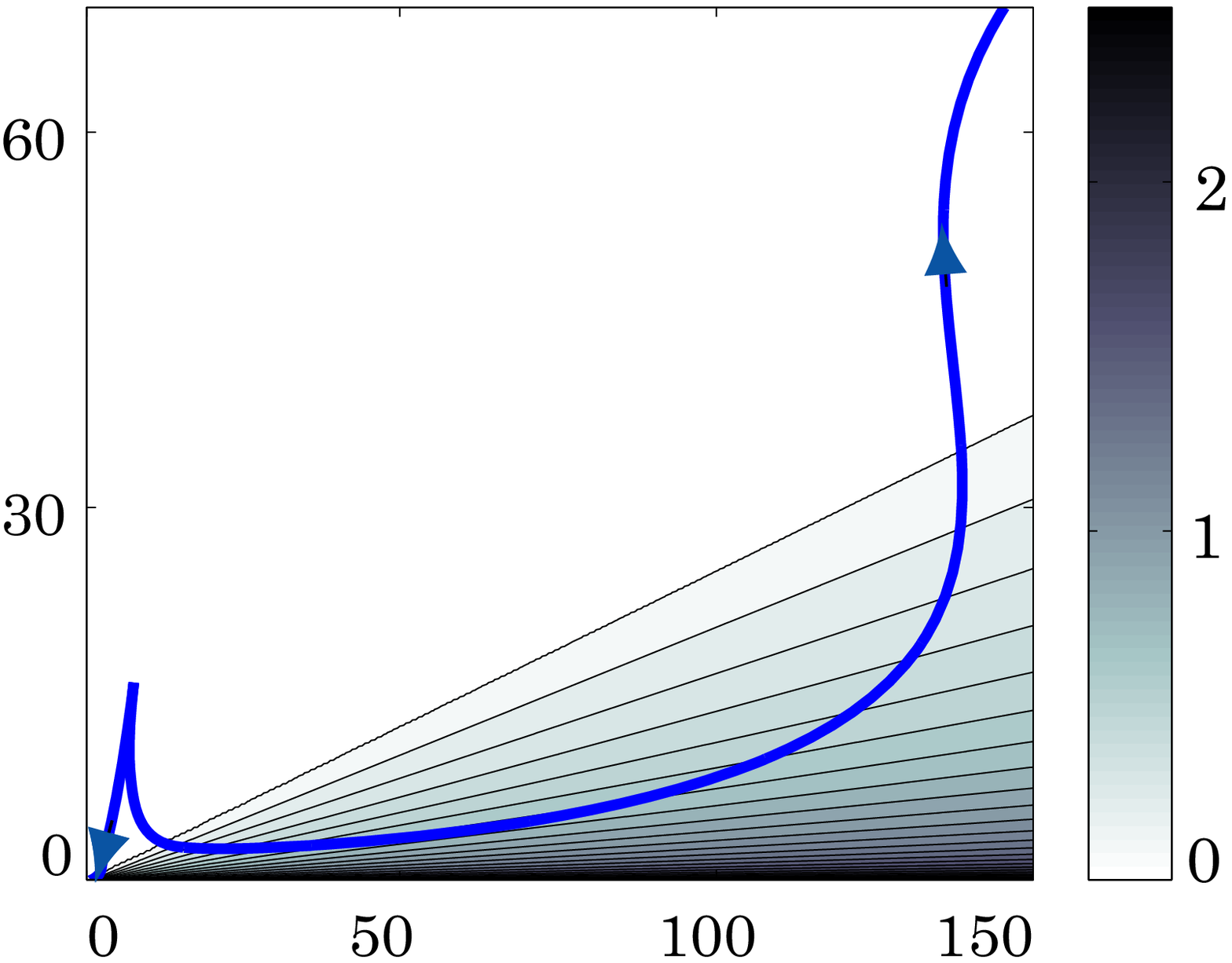}}
%\put(-21,-3)
%{\includegraphics[width=3.03in]{BigTraj2.png}}
\put(21,71)
{\includegraphics[width=1.05in]{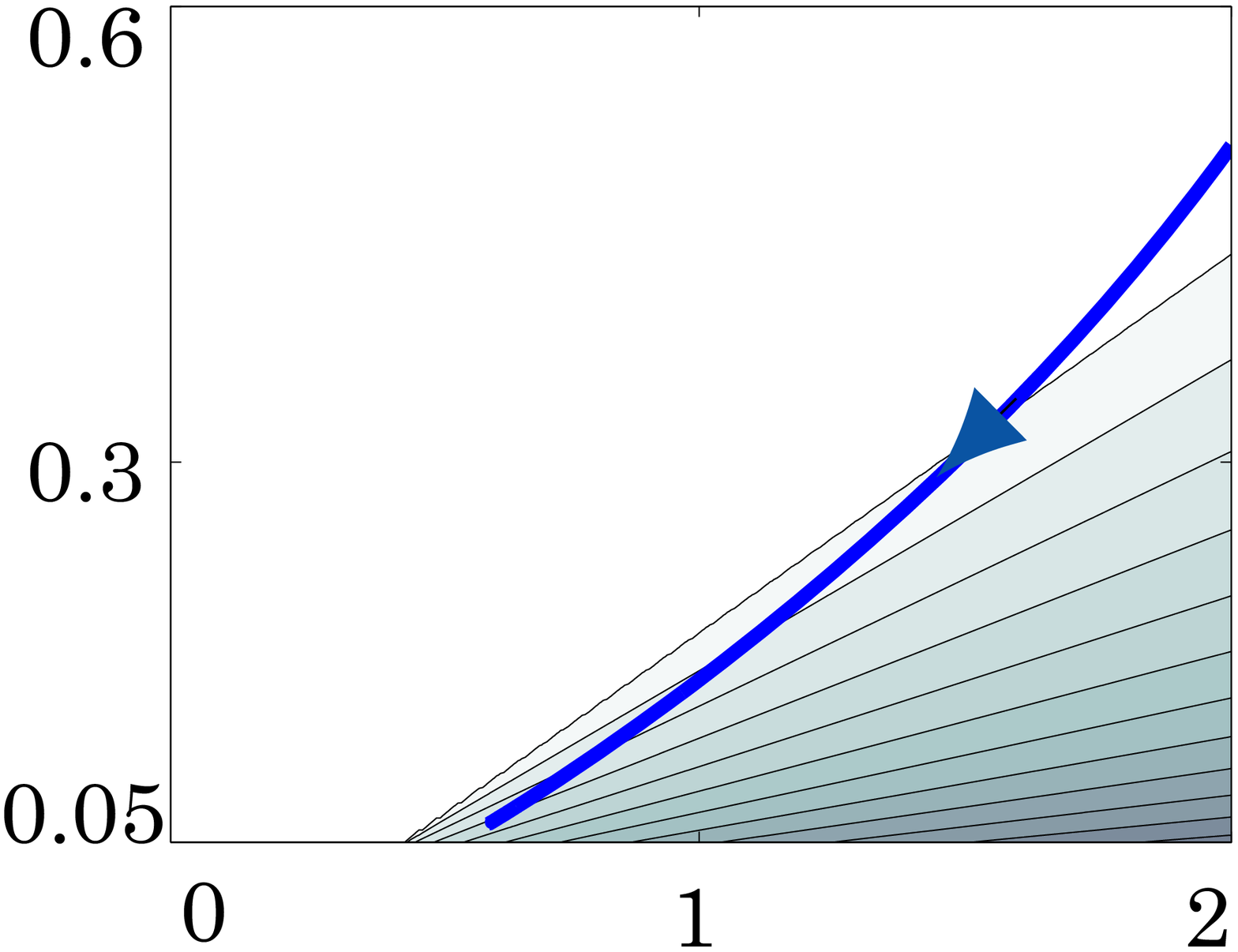}}
%\put(19.7,70.3)
%{\includegraphics[width=1.14in]{SmallTraj2.png}}
\put(10,16){\line(1,3){20.9}}
\put(10,16){\line(4,3){84}}
\put(73,1){$\tilde{\Delta}$}
\put(-22,99){${\rm det}(V)$}
\put(177,138){$E_{\cal N}$}
\put(21,43){$t=0$}
\put(108,87){$t=6.2$}
\put(43,65){$t=+\infty$}
\put(14,45){\circle{7}}
\put(50,80){\circle{7}}
\put(140.5,80.8){\circle{7}}
\put(48,107){${\cal T}_{\rm dam}$}
\put(70,94){\line(-1,1){10}}
\put(-14,40){${\cal T}_{\rm dam}$}
\put(11,27){\line(-1,1){10}}
\put(120,118){${\cal T}_{\rm dis}$}
\end{picture}

\caption{
Trajectories of the parameter $(\tilde{\Delta},{\rm det}(V))$
without feedback control. 
${\cal T}_{\rm dam}$ and ${\cal T}_{\rm dis}$ correspond to the
damped-damped and dispersive-damped cases, respectively.
We set $V_0=2I$ at $t=0$, from which $\tilde{\Delta}=8$ and
${\rm det}(V_0)=16$ follow.
($\tilde{\Delta}$ and ${\rm det}(V)$ are in units of $\hbar=1$ and 
$\hbar^2=1$, respectively.) 
 }
\end{figure}

%%%%%%%%%%%%%%%%%%%%%%%%%%%%%%%%%%%%%%%%%%%%%%%%%%%%%%%%%%%%%%%%%%%%%%%%%
%%%%%%%%%%%%%%%%%%%%%%%%%%% Feedback control %%%%%%%%%%%%%%%%%%%%%%%%%%%%
%%%%%%%%%%%%%%%%%%%%%%%%%%%%%%%%%%%%%%%%%%%%%%%%%%%%%%%%%%%%%%%%%%%%%%%%%

\subsection{Feedback control}

We first discuss how to determine the coefficient vector
$f=(f_1\trans, f_2\trans)\trans$ that realizes high-quality
entanglement control. Fortunately, in the ideal case, we can
explicitly find such an $f$. The idea was originally provided by
Wiseman and Doherty in \cite{Wiseman3}, but  here we apply the
idea in a slightly different manner.

Assume that $g=1$. Then, the Lyapunov equation \eqref{Lyapunov Eq}
with coefficient matrices $A_{\rm id}$ and $D_{\rm id}$ in Eqs.
\eqref{ideal controlled A} and \eqref{ideal controlled D}
can be rewritten as
\begin{equation}
\label{controlled lyapunov}
    \frac{dV_t}{dt}
    ={\cal R}(V_t)
       +\Sigma_2(f-f_t)(f-f_t)\trans\Sigma_2\trans,
\end{equation}
where
\begin{equation}
\label{v vector}
    f_t:=2\Sigma_2 V_t{\rm Re}(\ell)+{\rm Im}(\ell)
\end{equation}
and
\begin{eqnarray}
& & \hspace*{-1em}
\label{Riccati eq}
    {\cal R}(V):=A_o V+VA_o\trans+D_o
\nonumber \\ & & \hspace*{2em}
    \mbox{}
       -\Big[2V{\rm Re}(\ell)-\Sigma_2{\rm Im}(\ell)\Big]
                   \Big[2V{\rm Re}(\ell)-\Sigma_2{\rm Im}(\ell)\Big]\trans.
\nonumber
\end{eqnarray}
We now recall from Fig.~2~(b) that entanglement sudden-death
occurs simultaneously with a decrease of the purity
\eqref{purity}. This suggests that preventing a decrease of purity
may also prevent  entanglement sudden-death. However, we should point
out that it is not apparent that this will always be the case
and the relationship between loss of purity and entanglement
sudden-death needs to be studied further. Therefore, a simple
control strategy that we try here is to find a feedback controller
that prevents an increase of ${\rm det}(V_t)$ in order to keep the
purity high. As the second term of the right-hand side of Eq.
\eqref{controlled lyapunov} is always non-negative, it is then
reasonable to choose the time-variant coefficient vector
$f=f_t$. Of course, this intuitive argument does not always
allow us to conclude that ${\rm det}(V_t)$ takes its minimum
value. However, it is known that the {\it algebraic Riccati
equation} ${\cal R}(V)=0$ has a solution satisfying 
${\rm det}(V)=1/16$, which implies that the maximum purity $P=1$ is
achieved; e.g., see \cite{Wiseman3}.  Now assume that Eq.
\eqref{controlled lyapunov} has a unique steady solution
$V_\infty$ for a constant $f$. Then, by taking the time-invariant
coefficient vector
\begin{equation}
\label{best f}
   \bar{f}:=2\Sigma_2 V_\infty{\rm Re}(\ell)+{\rm Im}(\ell),~~~
        {\cal R}(V_\infty)=0,
\end{equation}
we obtain the same desirable result, ${\rm det}(V_\infty)=1/16$.
Note that the numerical solution to the algebraic Riccati equation
${\cal R}(V_\infty)=0$ can readily be obtained using a standard
software package such as {\sc matlab}.

We now consider direct feedback control with the coefficient
vector \eqref{best f}. Let us begin with the case where the first
cavity-field interaction occurs dispersively. For this system, it
is expected from Fig.~3 that the trajectory ${\cal T}_{\rm dis}$
can be modified and stabilized via feedback so that it has an
equilibrium point in the area where $E_{\cal N}$ is large. That
this is indeed true is shown below. When the parameters are given
by $m=0.2$ and $\kappa=1$, we find that
$\bar{f}=(0.1212,2.2196,-0.3163,-3.2277)\trans$ from (\ref{best
  f}). Fig.~5
illustrates that the controlled trajectory, denoted by ${\cal
T}_{\rm dis}^{\rm c}$, indeed shows the convergence that we had
hoped for. The entanglement and the purity of the steady cavity
state are shown in Fig.~4. While $\bar{f}$ is determined with
fixed $g=1$, we consider variations in $g$ to gain understanding
about its effect on the control system. When control is not used
($g=0$), the pair of dispersive and damped cavities does not
settle down to a steady state as seen in Section~III-A, and we
find that $E_{\cal N}\rightarrow 0$ as $t\rightarrow\infty$. On the other
hand, even with the small-gain feedback controller, the system
becomes stable and has a unique steady state with nonzero
entanglement. Remarkably, when $g=1$, the entanglement of the
steady state ($E_{\cal N}\approx 2.2$) improves upon the maximum
value of $E_{\cal N}$ of the uncontrolled state ($E_{\cal
N}\approx 0.65$) shown in Fig.~2~(a). Hence we see that  direct
feedback not only prevents entanglement sudden-death, but can also
enhance the entanglement.

Feedback can also improve the entanglement of a system where both
cavities are damped, but it is still very small as seen from the
dotted line in Fig.~4~(a). (The coefficient vector defined by Eq.
\eqref{best f} in this case is calculated to be
$\bar{f}=(0.0629,0.1525,0.2479,-0.5830)$.) To understand this
phenomenon, we recall that the uncontrolled trajectory ${\cal
T}_{\rm dam}$ has an equilibrium point that is located far from
the area with large $E_{\cal N}$. Hence, it should be hard to
drastically modify this trajectory such that it could reach that
area.
It is actually observed in Fig.~5 that, even with the vector
$\bar{f}$, the controlled trajectory ${\cal T}_{\rm dam}^{\rm c}$
shows almost the same time-evolution as the autonomous one
${\cal T}_{\rm
dam}$.

The above results suggest that strong stability of the autonomous
system sometimes makes it difficult for the state to transit into
a desirable entangled target.

\begin{figure}[t]
\centering

\begin{picture}(170,145)
\put(1,1)
{\includegraphics[width=2.4in]{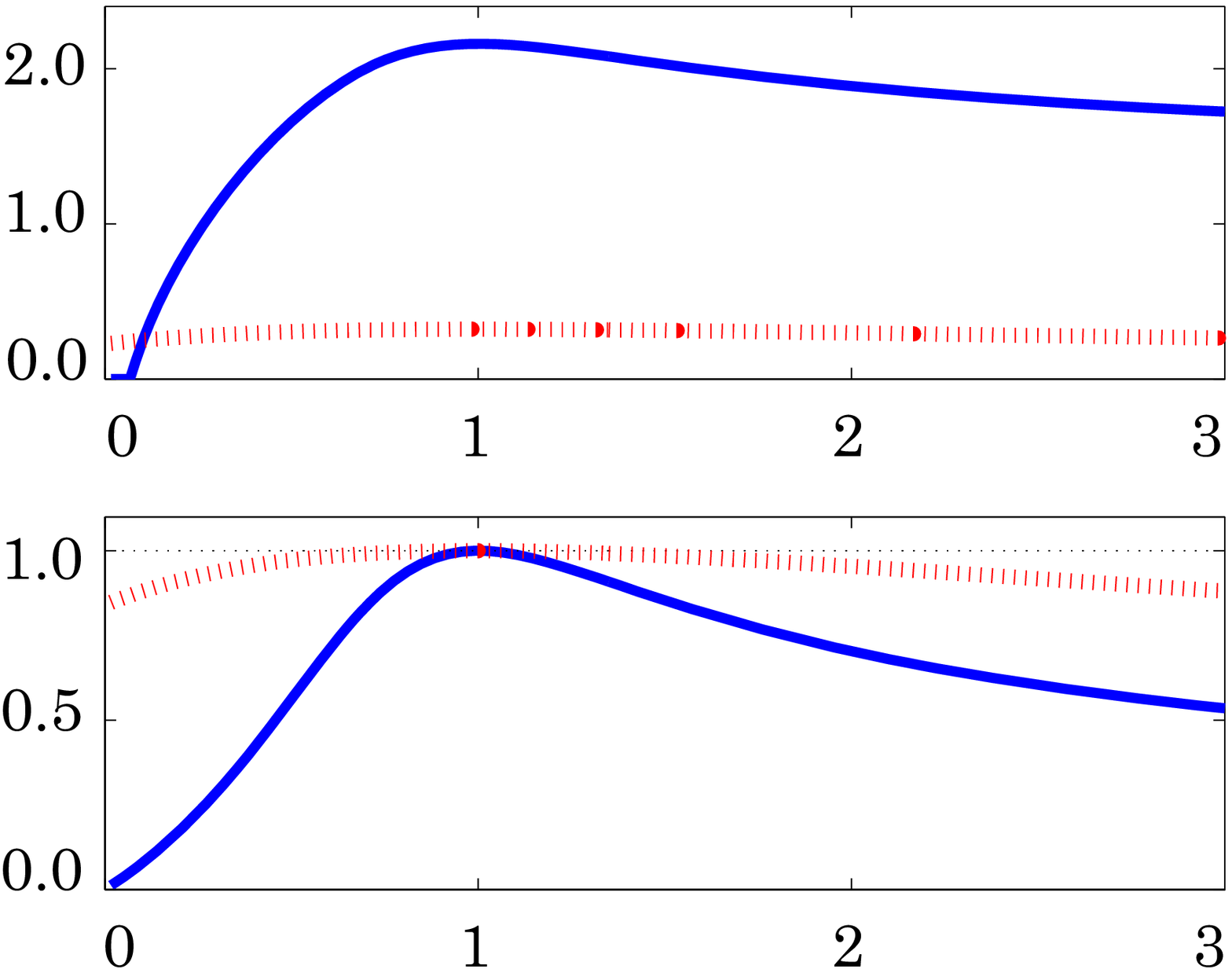}}
%\put(-4.5,0)
%{\includegraphics[width=2.65in]{ControlSudden.png}}
\put(91,2){$g$}
\put(91,72){$g$}
\put(-12,114){$E_{\cal N}$}
\put(-11,45){$P$}
\put(172,123){(a)}
\put(172,54){(b)}
\end{picture}

\caption{
The logarithmic negativity (a) and the purity (b) of the steady
cavity state with feedback control.
$g$ is the control gain.
The solid and dotted lines correspond to the dispersive-damped
and damped-damped cases, respectively.
The parameters are $m=0.2$ and $\kappa=1$.
}
\end{figure}

\begin{figure}[t]
\centering

\begin{picture}(170,155)
\put(1,1)
{\includegraphics[width=2.6in]{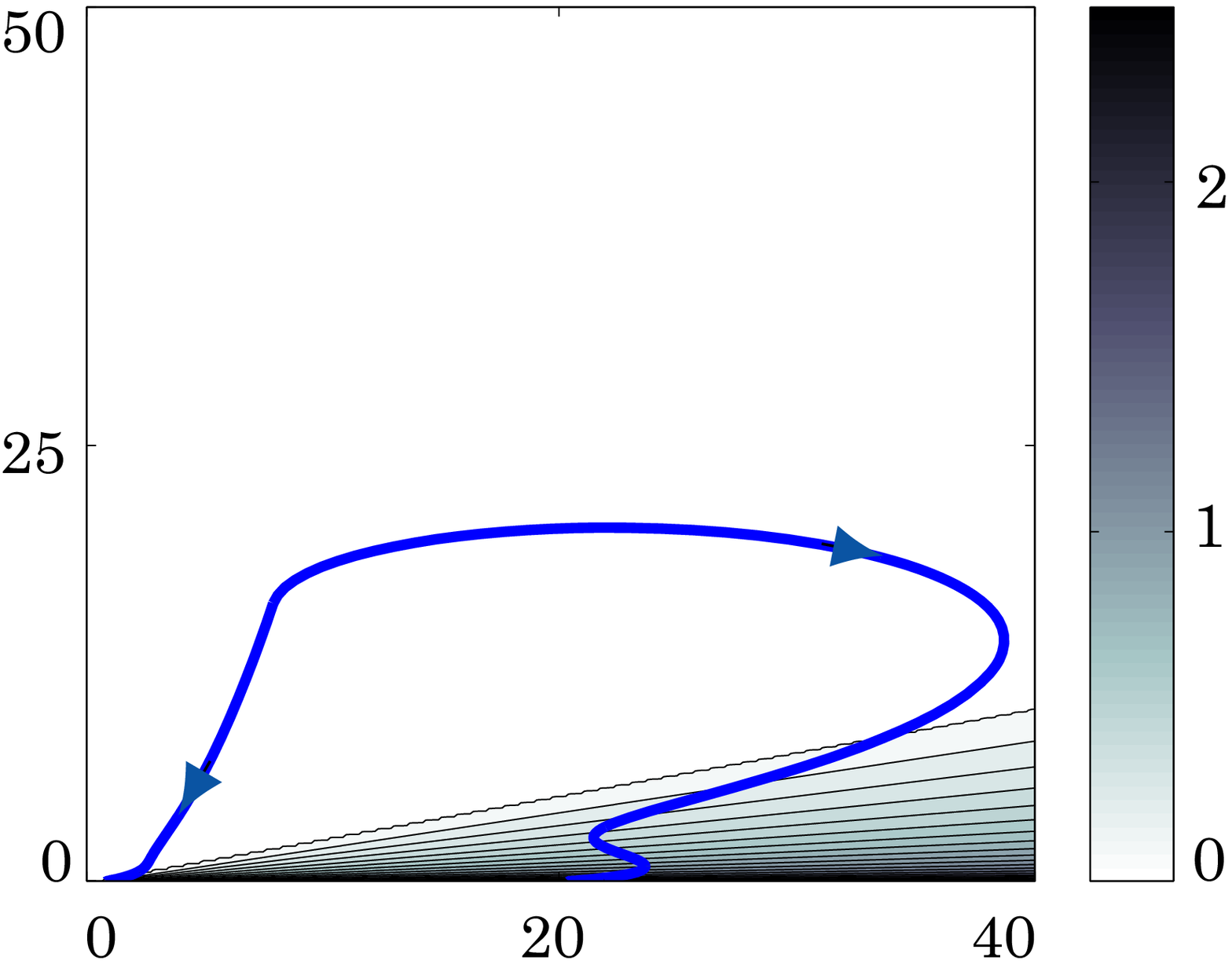}}
%\put(-13,-3)
%{\includegraphics[width=3.03in]{ControlBigTraj2.png}}
\put(42,79)
{\includegraphics[width=1.05in]{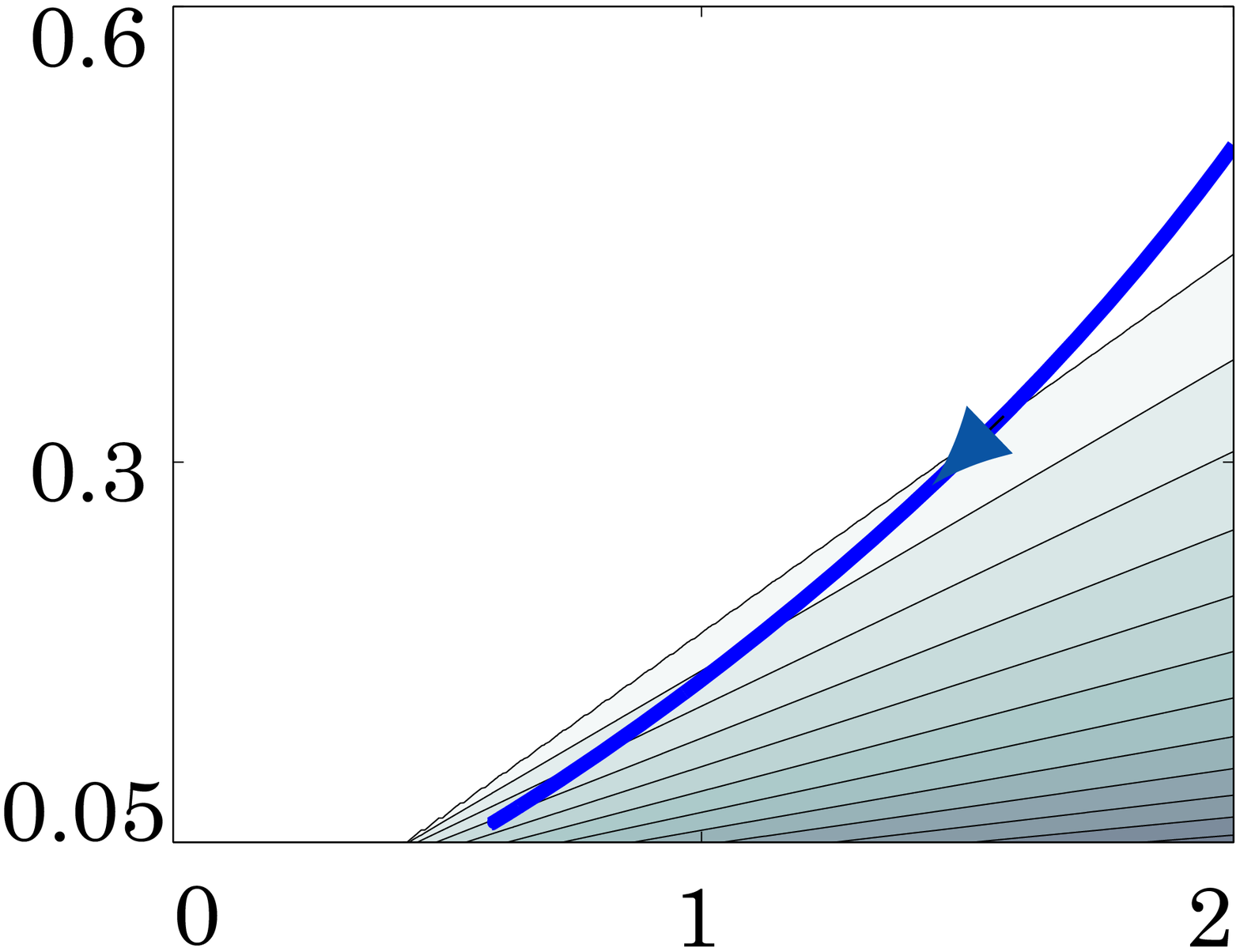}}
%\put(41.5,78)
%{\includegraphics[width=1.14in]{ControlSmallTraj2.png}}
\put(17,16){\line(1,2){35.5}}
\put(21,16){\line(4,3){94}}
\put(119,0){$\tilde{\Delta}$}
\put(-16,112){${\rm det}(V)$}
\put(185,137){$E_{\cal N}$}
\put(48,53){$t=0$}
\put(108,47){$t=6.2$}
\put(63,74){$t=+\infty$}
\put(44,59){\circle{7}}
\put(71,88){\circle{7}}
\put(138,39){\circle{7}}
\put(69,115){${\cal T}_{\rm dam}^{\rm c}$}
\put(89,101){\line(-1,1){10}}
\put(16,57){${\cal T}_{\rm dam}^{\rm c}$}
\put(36,42){\line(-1,1){10}}
\put(135,70){${\cal T}_{\rm dis}^{\rm c}$}
\end{picture}

\caption{
Trajectories of the parameters $(\tilde{\Delta},{\rm det}(V))$
with feedback control.
${\cal T}_{\rm dam}^{\rm c}$ and ${\cal T}_{\rm dis}^{\rm c}$
correspond to the damped-damped and dispersive-damped cases,
respectively.
The initial state is the same as before: $V_0=2I$.}
\end{figure}

%%%%%%%%%%%%%%%%%%%%%%%%%%%%%%%%%%%%%%%%%%%%%%%%%%%%%%%%%%%%%%%%%%%%%%%%%
%%%%%%%%%%%%%%%%%%%%%%%%%%% The realistic model %%%%%%%%%%%%%%%%%%%%%%%%%
%%%%%%%%%%%%%%%%%%%%%%%%%%%%%%%%%%%%%%%%%%%%%%%%%%%%%%%%%%%%%%%%%%%%%%%%%

\section{A realistic control scenario}
\label{sec:real-entg-cntl}

Finally, we return to the original setup of the network.
That is, the quantum channel is in contact with an environment,
and the homodyne detector is replaced by a realistic LPF with finite
bandwidth.
The purpose here is to study the impacts of these realistic
components on the entanglement of the cavity state.
The covariance matrix of the cavity state corresponds to
the left-upper $4\times 4$ submatrix of $V_t$ that is the
solution of
Eq. \eqref{Lyapunov Eq} with $A_{\rm re}$ and $D_{\rm re}$
given in Eqs. \eqref{A matrix} and \eqref{D matrix}.
Note that the cavity state is the reduced one with the detector
mode traced out.
We here focus only on the network where the first cavity interacts
with the field dispersively.
For the control Hamiltonian, we use the same coefficient
vector $\bar{f}=(0.1212,2.2196,-0.3163,-3.2277)\trans$.
It should be noted that, in this realistic case, we cannot
follow the  discussion in Section III-B to obtain a reasonable
coefficient vector $f$.

First, consider Fig.~6~(a).
This shows some plots of $E_{\cal N}$ with the time-constant
$\tau$ changing between $0.01\leq\tau\leq 0.6$ and with the
fixed transmittance $\alpha=1$ (i.e., no loss in the channel).
The most upper line almost coincides with the ideal one shown
in Fig.~4~(a).
That is, the entanglement in the realistic situation continuously
converges to the ideal one as $\tau\rightarrow 0$.
We also observe that the degradation of $E_{\cal N}$ is small
with respect to $\tau$.
Since the detector is regarded as a component of the controller,
these results imply that the realistic direct feedback
is robust against signal loss in the LPF.
In other words,  direct feedback control is reliable
even in this realistic situation.

On the other hand, Fig.~6~(b) plots $E_{\cal N}$ for some
values of the channel loss $\beta:=1-\alpha$ with fixed
$\tau=0.01$.
We find that $E_{\cal N}$ converges to the ideal one as
$\beta\rightarrow 0$, similar to the above case.
However, in this case, $E_{\cal N}$ rapidly decreases with
respect to $\beta$.
Even for the very small loss $\beta=0.01$, a visible
degradation occurs.
Moreover, when $\beta=0.1$, which still means we have a
high-quality quantum channel, $E_{\cal N}$ decreases less than
half of the ideal one.
That is, the entanglement is fragile to realistic channel
loss.

The above results are reasonable because the channel loss
directly reflects the decrease of interaction strength, while
the finite bandwidth of LPF simply implies loss of a classical
signal.
Hence the former should be a critical factor for entanglement
generation.

\begin{figure}[t]
\centering

\begin{picture}(170,145)
\put(8,1)
{\includegraphics[width=2.3in]{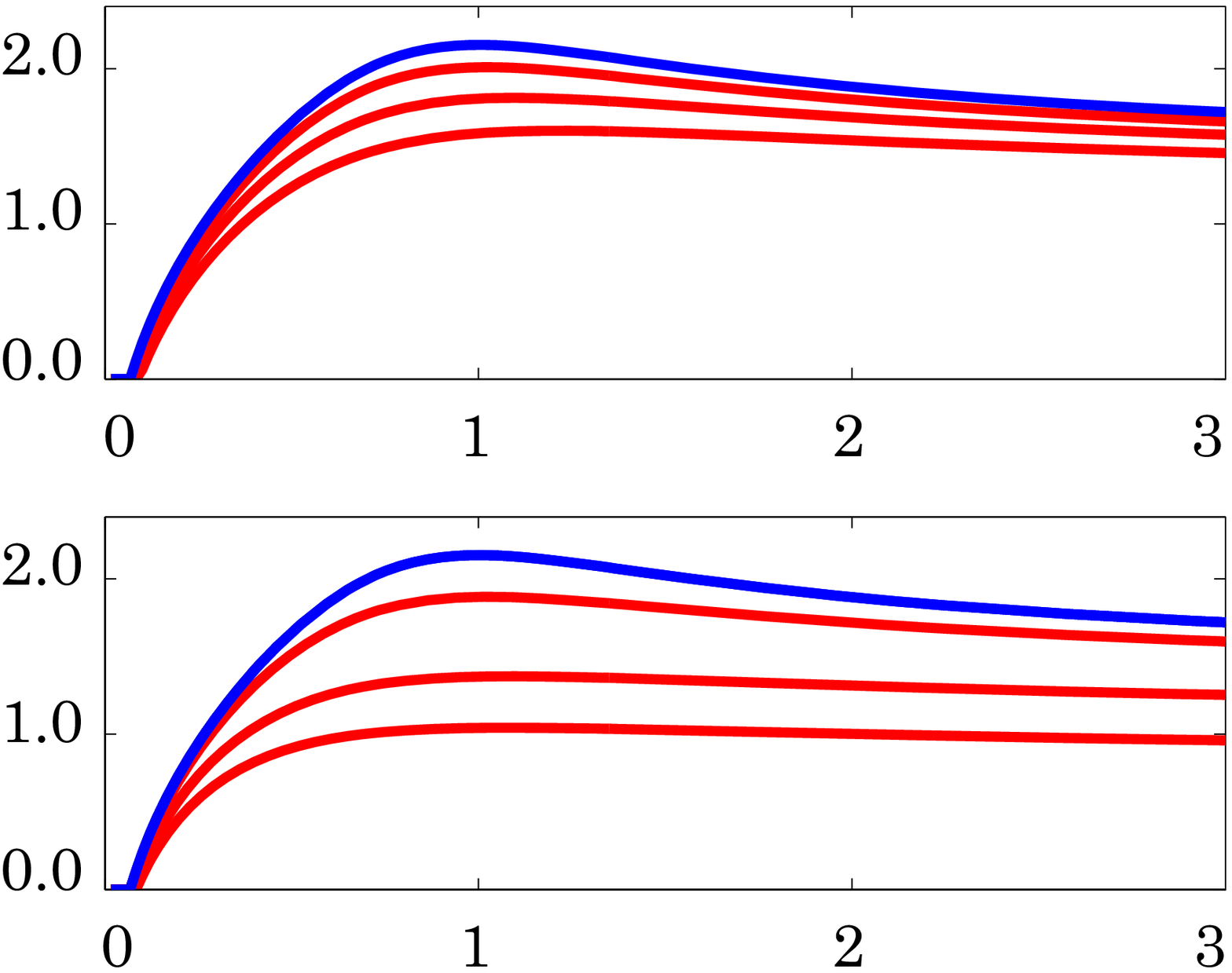}}
%\put(2,1)
%{\includegraphics[width=2.6in]{Entandegrade.png}}
\put(94,2){$g$}
\put(94,70){$g$}
\put(-11,111){$E_{\cal N}$}
\put(-11,43){$E_{\cal N}$}
\put(172,123){(a)}
\put(172,54){(b)}
\end{picture}

\caption{
The logarithmic negativity of the steady cavity state with
feedback control.
$g$ is the control gain.
(a) From the top downwards, the lines correspond to
$\tau = 0.01, 0.2, 0.4, 0.6$, while $\alpha=1$.
(b) From the top downwards, the lines correspond to
$\alpha=1,0.99,0.95,0.90$, while $\tau=0.01$.
In both cases, the LPF noise is very small; $a_4=0.01$.
The parameters are $m=0.2$ and $\kappa=1$.
}
\end{figure}

%%%%%%%%%%%%%%%%%%%%%%%%%%%%%%%%%%%%%%%%%%%%%%%%%%%%%%%%%%%%%%%%%%%%%%%%%
%%%%%%%%%%%%%%%%%%%%%%%%%%%%%% Conclusion %%%%%%%%%%%%%%%%%%%%%%%%%%%%%%%
%%%%%%%%%%%%%%%%%%%%%%%%%%%%%%%%%%%%%%%%%%%%%%%%%%%%%%%%%%%%%%%%%%%%%%%%%

\section{Conclusion}

The contributions of this paper are summarized as follows. First,
it was shown that, when the first cavity is undamped and the
second one is damped, the overall cavity state obtains a
significant amount of entanglement, which however disappears in a
finite time. Then, we have shown that direct measurement feedback
can avoid this entanglement sudden-death, and further, enhances
the entanglement. Moreover, it was shown that the direct feedback
controller is reliable under the influence of signal loss in a
realistic detector, although imperfection in the quantum channel
is a critical issue that largely degrades the achieved
entanglement. We believe that the case study we have presented provides useful insights that may be of use for more complex
 quantum network engineering.

%%%%%%%%%%%%%%%%%%%%%%%%%%%%%%%%%%%%%%%%%%%%%%%%%%%%%%%%%%%%%%%%%%%%%%%%%
%%%%%%%%%%%%%%%%%%%%%%%%%%%%%%%% Appendix %%%%%%%%%%%%%%%%%%%%%%%%%%%%%%%
%%%%%%%%%%%%%%%%%%%%%%%%%%%%%%%%%%%%%%%%%%%%%%%%%%%%%%%%%%%%%%%%%%%%%%%%%

\appendix

%%%%%%%%%%%%%%%%%%%%%%%%%%%%%%% Appendix A %%%%%%%%%%%%%%%%%%%%%%%%%%%%%%

\section{Quantum cascade systems}

\subsection{General theory}

In this appendix, we begin with a review of the theory of
quantum cascade systems which was originally developed
by Carmichael \cite{Carmichael1,Carmichael2} and Gardiner
\cite{Gardiner} in a quantum optics framework
and recently reformulated in more general setting by Gough
and James \cite{Gough}.
We then apply the theory to our model and derive the
corresponding system matrices.

The most general form of quantum dynamics that interacts
with $M$ optical input fields is described by the following
unitary evolution:
\begin{eqnarray}
& & \hspace*{-1em}
\label{general HP-eq}
   d\hat{U}_t
      =\Big[ \big(-\im \hat{H}
                  -\half\sum_{k=1}^{M}\hat{L}_k^* \hat{L}_k\big)dt
\nonumber \\ & & \hspace*{1.5em}
       \mbox{}
       +\sum_{k=1}^{M}\hat{L}_k d\hat{B}_{k,t}^*
       -\sum_{k,j=1}^{M}\hat{L}_j^* \hat{S}_{jk} d\hat{B}_{j,t}
\nonumber \\ & & \hspace*{1.5em}
       \mbox{}
       +\sum_{k,j=1}^{M}(\hat{S}_{kj}-\delta_{kj})d\hat{\Lambda}_{kj}
                        \Big]\hat{U}_t,~~~
   \hat{U}_0=\hat{I}.
\end{eqnarray}
This is also called the Hudson-Parthasarathy equation \cite{Hudson}.
Here, the operators $\hat{B}_{k,t}$ and $\hat{B}_{k,t}^*$ represent
the quantum annihilation and creation processes on the $k$-th field,
respectively.
The operator $\hat{\Lambda}_{kj}$ represents the scattering process
from the $k$-th state to the $j$-th state, and it satisfies
$d\hat{\Lambda}_{ij}d\hat{\Lambda}_{kl}=\delta_{jk}d\hat{\Lambda}_{il}$.
The matrix of operators $\hat{{\bf S}}:=(\hat{S}_{ij})$ must satisfy
$\hat{{\bf S}}\dgg\hat{{\bf S}}=\hat{{\bf S}}\hat{{\bf S}}\dgg=I$
in order for $\hat{U}_t$ to be unitary.
The system is completely characterized by the triple
${\bf \Gamma}=(\hat{{\bf S}},\hat{{\bf L}},\hat{H})$, where $\hat{{\bf L}}$
is a vector of operators
$\hat{{\bf L}}:=(\hat{L}_1,\ldots,\hat{L}_M)\trans$.
Let $\hat{X}$ be an operator of the system.
Then, this evolves in time according to the Heisenberg equation
$\hat{X}\rightarrow j_t(\hat{X}):=\hat{U}_t^*\hat{X}\hat{U}_t$.
In particular, we can define $M$ output fields
$\hat{B}'_{k,t}:=j_t(\hat{B}_{k,t})$, which yields
\begin{equation*}
\label{general output}
    d\hat{{\bf B}}'_t=j_t(\hat{{\bf L}})dt
                       +j_t(\hat{{\bf S}})d\hat{{\bf B}}_t,
\end{equation*}
where we have defined
$\hat{{\bf B}}'_t:=(\hat{B}'_{1,t},\ldots,\hat{B}'_{M,t})\trans,
j_t(\hat{{\bf L}}):=(j_t(\hat{L}_1),\ldots,j_t(\hat{L}_M))\trans$, etc.

Let us now consider two systems:
${\bf \Gamma}_1=(\hat{{\bf S}}_1,\hat{{\bf L}}_1,\hat{H}_1)$ and
${\bf \Gamma}_2=(\hat{{\bf S}}_2,\hat{{\bf L}}_2,\hat{H}_2)$.
Note that the number of inputs (i.e., outputs) of these systems
can always be matched by introducing additional components $0$ in
$\hat{{\bf L}}$ and $I$ in $\hat{{\bf S}}$ as $\hat{{\bf L}}\oplus 0$
and $\hat{{\bf S}}\oplus I$.
These systems can be connected so that the outputs of ${\bf \Gamma}_1$
are the inputs of ${\bf \Gamma}_2$, as  depicted abstractly in
Fig.~7.
We denote this cascade system by
${\bf \Gamma}_2\vartriangleleft{\bf \Gamma}_1$.
\begin{figure}[t]
\centering

\begin{picture}(220,100)
\put(45,2)
{\includegraphics[width=1.8in]{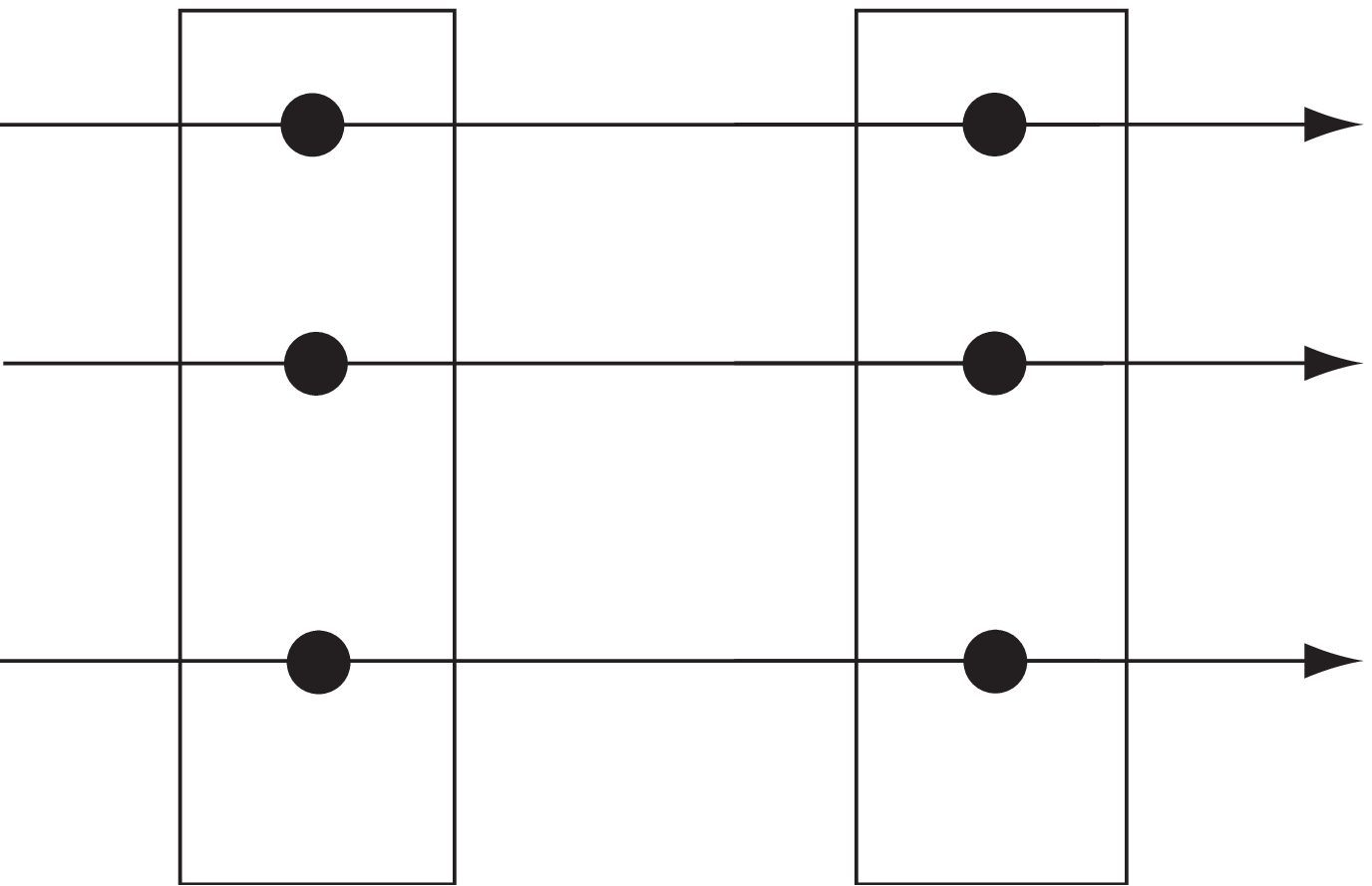}}
%\put(44.5,2)
%{\includegraphics[width=1.8in]{AbstGeneralCascade.png}}
\put(72,92){${\bf \Gamma}_1$}
\put(135,92){${\bf \Gamma}_2$}
\put(25,71){$\hat{B}_{1,t}$}
\put(25,50){$\hat{B}_{2,t}$}
\put(25,18){$\hat{B}_{M,t}$}
\put(99,62){$\hat{B}'_{1,t}$}
\put(99,40){$\hat{B}'_{2,t}$}
\put(99,10){$\hat{B}'_{M,t}$}
\put(67,61){$\hat{L}_{1,1}$}
\put(131,61){$\hat{L}_{1,2}$}
\put(67,38){$\hat{L}_{2,1}$}
\put(131,38){$\hat{L}_{2,2}$}
\put(65.5,9){$\hat{L}_{M,1}$}
\put(130,9){$\hat{L}_{M,2}$}
\end{picture}

\caption{
Abstract illustration of the cascade system.
The black circles represent that the subsystem
interacts with the field.
}
\end{figure}
Then, from \cite{Gough}, we have
\begin{eqnarray}
& & \hspace*{-1em}
\label{cascade formula}
    {\bf \Gamma}_2\vartriangleleft{\bf \Gamma}_1
     =\Big(\hat{{\bf S}}_2\hat{{\bf S}}_1,~
           \hat{{\bf L}}_2+\hat{{\bf S}}_2\hat{{\bf L}}_1,~
\nonumber \\ & & \hspace*{1.5em}
      \hat{H}_1+\hat{H}_2+
           \frac{1}{2\im}(
              \hat{{\bf L}}_2\dgg\hat{{\bf S}}_2\hat{{\bf L}}_1
             -\hat{{\bf L}}_1\dgg\hat{{\bf S}}_2\dgg\hat{{\bf L}}_2) \Big),
\end{eqnarray}
where $\hat{{\bf L}}_k\dgg=(\hat{L}_{1,k}^*,\ldots,\hat{L}_{M,k}^*),~
(k=1,2)$ and $\hat{{\bf S}}\dgg=(\hat{S}_{ji}^*)$.

Direct measurement feedback \cite{Wiseman1,Wiseman2} is no more
than a cascade of the system and the controller.
Hence, the overall system representation of the controlled
network is readily derived using Eq. \eqref{cascade formula}
as follows.
For simplicity, let us consider a single-input single-output system
${\bf \Gamma}=(\hat{S},\hat{L},\hat{H}+u_t\hat{F})$, where $u_t$
represents the control input.
An ideal homodyne detector yields a classical signal
$y_t=j_t(\hat{B}_t+\hat{B}_t^*)$.
Then, the direct feedback $u_t=gy_t$ closes the loop and realizes
\begin{eqnarray}
& & \hspace*{-1em}
\label{direct fb cascade}
    {\bf \Gamma}_{\rm fb}
      =(1,-\im g\hat{F},0)\vartriangleleft(\hat{S},\hat{L},\hat{H})
\nonumber \\ & & \hspace*{0.7em}
      =\Big(\hat{S},~\hat{L}-\im g\hat{F},~
        \hat{H}+\frac{g}{2}(\hat{F}\hat{L}+\hat{L}^*\hat{F}) \Big).
\end{eqnarray}
For a more detailed discussion, see \cite{Gough}.

\subsection{Ideal network}

Let us then apply the above formulas to our system. First, we
consider an ideal network composed of the following three
subsystems:
\begin{eqnarray}
& & \hspace*{-1em}
\label{three components}
   \mbox{Cavity 1:}~~
   {\bf C}_1=( I,~\hat{L}_{1,1},~\hat{H}_1 ),
\nonumber \\ & & \hspace*{-1em}
   \mbox{Cavity 2:}~~
   {\bf C}_2=( I,~\hat{L}_{1,2},~\hat{H}_2 ),
\nonumber \\ & & \hspace*{-1em}
   \mbox{Controller:}~~
   {\bf F}=( I,\hspace{0.3em} -\im g\hat{F},~0 ),
\nonumber
\end{eqnarray}
where $\hat{L}_{1,i}$ and $\hat{H}_i$ are given in Section~II, and
$\hat{F}:=\hat{F}_1+\hat{F}_2
=f_1\trans\hat{{\bf x}}_1+f_2\trans\hat{{\bf x}}_2$.
The abstract configuration of the network is given in Fig.~8.
\begin{figure}[t]
\centering

\begin{picture}(220,98)
\put(45,7)
{\includegraphics[width=1.8in]{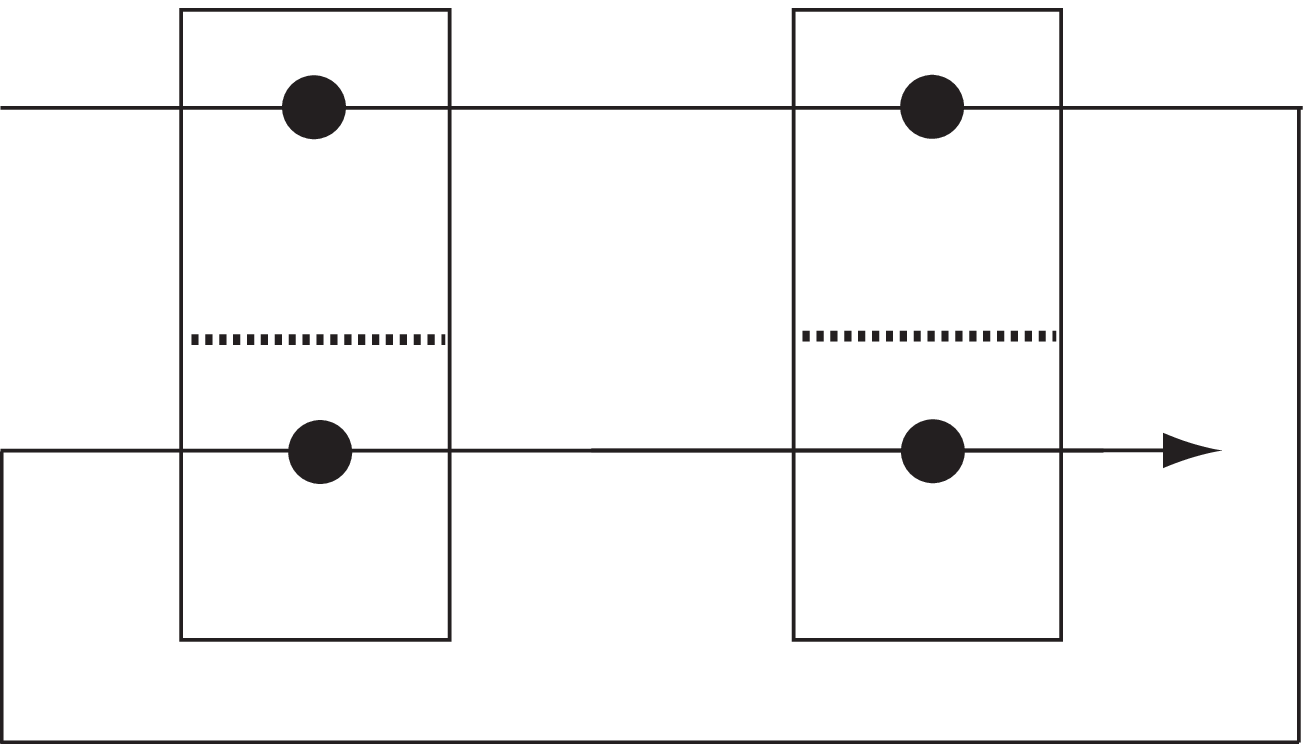}}
%\put(44.5,7)
%{\includegraphics[width=1.8in]{AbstIdealNetwork.png}}
\put(58,85){Cavity 1}
\put(119,85){Cavity 2}
\put(25,68){$\hat{B}_{1,t}$}
\put(69,55){$\hat{L}_{1,1}$}
\put(130.5,55){$\hat{L}_{1,2}$}
\put(124.5,23){$-\im g\hat{F}_2$}
\put(65,23){$-\im g\hat{F}_1$}
\end{picture}

\caption{
Abstract illustration of the ideal network.
}
\end{figure}
Thus, the network is given by
\begin{equation}
    {\bf F}\vartriangleleft{\bf C}_2
           \vartriangleleft{\bf C}_1
         =(I,~\hat{L}_{\rm id},~\hat{H}_{\rm id}),
\end{equation}
where
\begin{eqnarray}
& & \hspace*{-1em}
    \hat{L}_{\rm id}
      =\hat{L}_{1,1}+\hat{L}_{1,2}-\im g\hat{F}
\nonumber \\ & & \hspace*{0.7em}
      =[\ell_1\trans-\im g f_1\trans,~\ell_2\trans-\im g f_2\trans]
         \left[ \begin{array}{c}
          \hat{{\bf x}}_1 \\
          \hat{{\bf x}}_2 \\
         \end{array} \right]
\nonumber \\ & & \hspace*{0.7em}
      =:L_{\rm id}\hat{{\bf x}}
\nonumber
\end{eqnarray}
and
\begin{eqnarray}
& & \hspace*{-1em}
   \hat{H}_{\rm id}
         =\hat{H}_1+\hat{H}_2
           +\frac{1}{2\im}(\hat{L}_{1,2}^*\hat{L}_{1,1}
                       -\hat{L}_{1,1}^*\hat{L}_{1,2})
\nonumber \\ & & \hspace*{5em}
    \mbox{}
           +\frac{g}{2}\Big[\hat{F}(\hat{L}_{1,1}+\hat{L}_{1,2})
                          +(\hat{L}_{1,1}^*+\hat{L}_{1,2}^*)\hat{F}\Big]
\nonumber \\ & & \hspace*{0.7em}
    =\half\hat{{\bf x}}\trans
       \Big\{
         \left[ \begin{array}{cc}
          G_1 & {\rm Im}(\ell_1\ell_2\dgg) \\
          {\rm Im}(\ell_1\ell_2\dgg)\trans & G_2
         \end{array} \right]
\nonumber \\ & & \hspace*{5em}
      \mbox{}
        +gf{\rm Re}(\ell)\trans+g{\rm Re}(\ell)f\trans\Big\}\hat{{\bf x}}
\nonumber \\ & & \hspace*{0.7em}
    =:\half\hat{{\bf x}}\trans G_{\rm id}\hat{{\bf x}}.
\nonumber
\end{eqnarray}
Here, $\ell:=(\ell_1\trans, \ell_2\trans)\trans$ and
$f:=(f_1\trans, f_2\trans)\trans$.
From the definition, we then obtain the system $A$ and $D$ matrices:
$A_{\rm id}
:=\Sigma_2[G_{\rm id}+{\rm Im}(L_{\rm id}\dgg L_{\rm id})]$ and
$D_{\rm id}
:=\Sigma_2{\rm Re}(L_{\rm id}\dgg L_{\rm id})\Sigma_2\trans$,
and they are given in Eqs. \eqref{ideal controlled A} and
\eqref{ideal controlled D}.

\subsection{Realistic network}

We next consider a realistic model of the network.
Each component is given as follows.
\begin{eqnarray}
& & \hspace*{-1em}
\label{five components}
   \mbox{Cavity 1:}~~
   {\bf C}_1=\Big(I,~
              \left[ \begin{array}{c}
                \hat{L}_{1,1} \\
                0 \\
                0 \\
              \end{array} \right],~
              \hat{H}_1\Big),
\nonumber \\ & & \hspace*{-1em}
   \mbox{Beam Splitter:}~~
   {\bf B}=\Big(\left[ \begin{array}{ccc}
                \alpha & -\beta & 0 \\
                \beta & \alpha & 0 \\
                0 & 0 & 1 \\
             \end{array} \right],~ 0,~ 0\Big),
\nonumber \\ & & \hspace*{-1em}
   \mbox{Cavity 2:}~~
   {\bf C}_2=\Big(I,~
              \left[ \begin{array}{c}
                \hat{L}_{1,2} \\
                0 \\
                0 \\
              \end{array} \right],~
              \hat{H}_2\Big),
\nonumber \\ & & \hspace*{-1em}
   \mbox{Detector:}~~
   {\bf D}=\Big(I,~
              \left[ \begin{array}{c}
                \hat{L}_{1,3} \\
                0 \\
                \hat{L}_{3,3} \\
              \end{array} \right],~
              \hat{H}_3\Big),
\nonumber \\ & & \hspace*{-1em}
   \mbox{Controller:}~~
   {\bf F}=\Big(I,~
              \left[ \begin{array}{c}
                0 \\
                0 \\
                -\im g\hat{F} \\
              \end{array} \right],~
              0 \Big).
\nonumber
\end{eqnarray}
The $k$-th element of the above vectors corresponds to the $k$-th
quantum field $\hat{B}_{k,t}$.
Fig.~9 abstractly illustrates the structure of the interactions
in the network.
Note that the detector part includes the beam splitter with local
oscillator shown in Fig.~1.
\begin{figure}[t]
\centering

\begin{picture}(220,124)
\put(1,7)
{\includegraphics[width=3.1in]{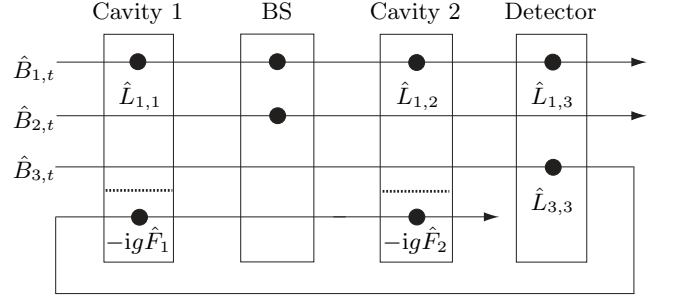}}
%\put(1,7)
%{\includegraphics[width=3.1in]{AbstRealNetwork.png}}
\put(15,111){Cavity 1}
\put(79,111){BS}
\put(120,111){Cavity 2}
\put(171,111){Detector}
\put(-15,51){$\hat{B}_{3,t}$}
\put(-15,70){$\hat{B}_{2,t}$}
\put(-15,89){$\hat{B}_{1,t}$}
\put(25,79){$\hat{L}_{1,1}$}
\put(130,79){$\hat{L}_{1,2}$}
\put(181,79){$\hat{L}_{1,3}$}
\put(181,39){$\hat{L}_{3,3}$}
\put(125,24){$-\im g\hat{F}_2$}
\put(20,24){$-\im g\hat{F}_1$}
\end{picture}

\caption{
Abstract illustration of the realistic network.
}
\end{figure}
Iteratively using Eq. \eqref{cascade formula}, we obtain
\begin{eqnarray}
& & \hspace*{-1em}
    {\bf F}\vartriangleleft{\bf D}
           \vartriangleleft{\bf C}_2
           \vartriangleleft{\bf B}
           \vartriangleleft{\bf C}_1
\nonumber \\ & & \hspace*{0em}
    =\Big(\left[ \begin{array}{ccc}
              \alpha & -\beta & 0 \\
              \beta & \alpha & 0 \\
              0 & 0 & 1 \\
          \end{array} \right],~
          \left[ \begin{array}{c}
                \alpha\hat{L}_{1,1}+\hat{L}_{1,2}+\hat{L}_{1,3} \\
                \beta\hat{L}_1 \\
                \hat{L}_{3,3}-\im g\hat{F} \\
          \end{array} \right],~\hat{H}_{\rm re}\Big),
\nonumber
\end{eqnarray}
where
\begin{eqnarray}
& & \hspace*{-1em}
     \hat{H}_{\rm re}=
       \hat{H}_1+\hat{H}_2+\hat{H}_3
           +\frac{\alpha}{2\im}(\hat{L}_{1,2}^*\hat{L}_{1,1}
                       -\hat{L}_{1,1}^*\hat{L}_{1,2})
\nonumber \\ & & \hspace*{2.7em}
       \mbox{}
       +\frac{1}{2\im}\Big[\hat{L}_{1,3}^*(\alpha\hat{L}_{1,1}+\hat{L}_{1,2})
             -(\alpha\hat{L}_{1,1}^*+\hat{L}_{1,2}^*)\hat{L}_{1,3}\Big]
\nonumber \\ & & \hspace*{2.7em}
       \mbox{}
           +\frac{g a_4}{2}(\hat{F}\hat{L}_{3,3}+\hat{L}_{3,3}^*\hat{F}).
\nonumber
\end{eqnarray}
It should be noted that $a_4$ appears in the last term of
$\hat{H}_{\rm re}$ due to $d\hat{B}_{3,t}d\hat{B}_{3,t}^*=a_4 dt$.
We now look at the relation
$(\hat{{\bf S}},\hat{{\bf L}},\hat{H})=(\hat{{\bf S}},0,0)\vartriangleleft
(I,\hat{{\bf S}}\dgg\hat{{\bf L}},\hat{H})$.
This implies that any system $(\hat{{\bf S}},\hat{{\bf L}},\hat{H})$
is equivalent to the system without scattering noises,
$(I,\hat{{\bf S}}\dgg\hat{{\bf L}},\hat{H})$, as long as we focus
on the unconditional state.
This is because $(\hat{{\bf S}},0,0)$ just
corresponds to the modulation of the output that is to be discarded.
(Note that if we consider the conditional state based on
the measurement result, the above equivalence does not hold.)
Consequently, our system is given by
\begin{equation}
   {\bf \Gamma}=(I,~\hat{{\bf L}}_{\rm re},~\hat{H}_{\rm re}),
\end{equation}
where
\begin{eqnarray}
\label{general L}
& & \hspace*{-1em}
   \hat{{\bf L}}_{\rm re}
         =\left[ \begin{array}{c}
             \hat{L}_{1,1}+\alpha\hat{L}_{1,2}+\alpha\hat{L}_{1,3} \\
             -\beta\hat{L}_{1,2}-\beta\hat{L}_{1,3} \\
             \hat{L}_{3,3}-\im g\hat{F} \\
          \end{array} \right]
\nonumber \\ & & \hspace*{0.7em}
         =\left[ \begin{array}{cccc}
             \ell_1\trans & \alpha\ell_2\trans & 0 & -\im \alpha a_2 \\
             0            & -\beta\ell_2\trans & 0 & \im \beta a_2 \\
             -\im gf_1\trans & -\im gf_2\trans       & a_3/2a_4 & 0 \\
          \end{array} \right]
          \left[ \begin{array}{c}
            \hat{{\bf x}}_1 \\
            \hat{{\bf x}}_2 \\
            \hat{{\bf x}}_3 \\
          \end{array} \right]
\nonumber \\ & & \hspace*{0.7em}
         =:L_{\rm re}\hat{{\bf x}}
\end{eqnarray}
and $\hat{H}_{\rm re}=\hat{{\bf x}}\trans G_{\rm re}\hat{{\bf x}}/2$ with
\begin{equation}
\label{general G}
    G_{\rm re}
     =\left[ \begin{array}{cccc}
        G_1 & \alpha{\rm Im}(\ell_1\ell_2\dgg) & a_3 g f_1/2
                                        & \alpha a_2{\rm Re}(\ell_1) \\
        \alpha{\rm Im}(\ell_1\ell_2\dgg)^{{\mathsf T}} & G_2
                         & a_3 g f_2/2 & a_2{\rm Re}(\ell_2) \\
        a_3 g f_1\trans/2 & a_3 g f_2\trans/2 & 0 & a_1 \\
        \alpha a_2 {\rm Re}(\ell_1)\trans & a_2{\rm Re}(\ell_2)\trans &
                                              a_1 & 0 \\
      \end{array} \right].
\end{equation}
Hence, we can now obtain the drift and diffusion matrices:
\begin{eqnarray}
& & \hspace*{-1em}
    \tilde{A}_{\rm re}
           =\Sigma_3[G_{\rm re}+{\rm Im}(L_{\rm re}\dgg L_{\rm re})]
           =\left[ \begin{array}{cc}
               A_{\rm re} & 0 \\
               0 & -a_1 \\
            \end{array} \right],
\nonumber \\ & & \hspace*{-1em}
    \tilde{D}_{\rm re}
           =\Sigma_3{\rm Re}(L_{\rm re}\dgg L_{\rm re})\Sigma_3\trans
           =\left[ \begin{array}{cc}
               D_{\rm re} & 0 \\
               0 & a_3^2/4a_4 \\
            \end{array} \right],
\end{eqnarray}
where $A_{\rm re}$ and $D_{\rm re}$ are given in Eqs. \eqref{A matrix}
and \eqref{D matrix}.
Since the $6$-th column and low elements do not affect the others,
it suffices to consider the reduced $5\times 5$ matrices
$A_{\rm re}$ and $D_{\rm re}$.

%%%%%%%%%%%%%%%%%%%%%%%%%%%%%%% Appendix B %%%%%%%%%%%%%%%%%%%%%%%%%%%%%%

\section{Realizations of linear stochastic models}

The purpose of this section is to discuss possible physical
realizations of the linear stochastic models making up the
entanglement scheme considered in Sections \ref{sec:entg-cntl} and
\ref{sec:real-entg-cntl}. Linear models are approximations of real
physical system that are valid under various assumptions such as
the dipole moment approximation and rotating wave approximation.
In particular, we will describe approximate realizations of these
models using optical cavities. Discussions of the physical meaning
of the abstract linear models considered herein can be found in,
e.g., \cite{Edwards,JNP06,NJD08}.

\subsection{Quadratic Hamiltonian}

Let $\hat{a}=(\hat{q}+\im\hat{p})/\sqrt{2}$ and
$\hat{a}^*=(\hat{q}-\im\hat{p})/\sqrt{2}$ be the cavity
annihilation and creation operators. Then, a quadratic Hamiltonian
\[
   \hat{H}
      =\Delta \hat{a}^*\hat{a}
        +\frac{\im}{2}(\epsilon{\rm e}^{\im\phi}\hat a^*\mbox{}^2
                       -\epsilon^*{\rm e}^{-\im\phi}\hat a^2)
\]
can be realized with a degenerate parametric amplifier (DPA) with
a classical pump \cite[Section 10.2]{Gardiner} in a rotating frame
at half the pump frequency, where $\epsilon e^{\im \phi}$
($\epsilon,\phi$ real) is the effective pump intensity and
$\Delta$ is the detuning frequency of the cavity mode of the DPA
from the half the pump frequency (i.e.,
$\Delta=\omega_{cav}-\omega_p/2$, where $\omega_{cav}$ is the
cavity resonance frequency and $\omega_p$ is the pump frequency).
It is then easy to verify that by choosing
\[
    \Delta=\frac{1+m}{2},~~\epsilon=\frac{1-m}{2},~~\phi=0
\]
the Hamiltonian can be written, in terms of the quadratures, as
$\hat{H}=(m\hat{q}^2+\hat{p}^2)/2$. The latter is the form of the
Hamiltonian used in our models.

%%%%%%%%%%%%%%%%%%%%%%%%%%%%%%%%%%%%%%%%%%%%%%%%%%%%%%%%%%%%%%%%%%%%%%

\subsection{Models with dissipative coupling
$\hat L=\sqrt{\kappa}\hat a$ and direct measurement feedback}

\begin{figure}
\includegraphics[scale=0.6]{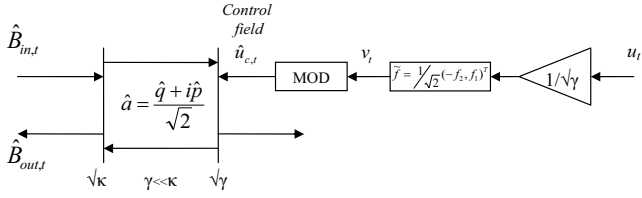}
\caption{ \label{fg_cav_diss} Implementation of a dissipative
coupling together with the linear control. }
\end{figure}

Linear systems with the dissipative coupling $\hat L=\sqrt{\kappa}
\hat a$ are quite standard and can be implemented as an optical
cavity with a leaky mirror, but here we shall also consider how
the direct measurement feedback term $\hat F=u_t f^{\top} \hat
{\bf x}$, with $f=(f_1,f_2)\trans$, can be implemented in this
system. Such an implementation is shown in Fig.~\ref{fg_cav_diss}.
The cavity has two partially transmitting mirrors $M_1$ and $M_2$
with coupling constants $\sqrt{\kappa}$ and $\sqrt{\gamma}$,
respectively. Here $\gamma$ is chosen such that $\gamma \ll 1$ and
$\gamma \ll \kappa$. The cavity interacts with an incident vacuum
noise field at mirror $M_1$ via the dissipation coupling $\hat
L=\sqrt{\kappa}\hat a$. The feedback $\hat F$ is implemented as
follows. First, the (real-valued) control signal $u_t$ is
amplified with gain $1/\sqrt{\gamma}$ and multiplied with $\tilde
f=\frac{1}{\sqrt{2}}(-f_2,f_1)\trans$ to give the real signal
$v_t=(v_{1,t},v_{2,t})\trans=\frac{1}{\sqrt{\gamma}}\tilde f u_t$.
$v_t$ is then sent to a modulator that displaces a vacuum bosonic
field by the classical field $\int_{0}^{t}v_s^C ds$ with
$v_t^C=v_{1,t}+\im v_{2,t}$ to produce a coherent control field
$\hat u_{c,t}$ \cite{Edwards} satisfying
$d\hat u_{c,t}= v_t^Cdt + d\hat B_{o,t}$, where $\hat B_{o,t}$
is a vacuum noise field independent of $\hat B_{in,t}$.
This displacement can be physically implemented by an
electro-optical modulator, see \cite[Section III-B.5]{DHJ06}.
Mathematically, the displacement of a vacuum field $\hat B_{o,t}$
by a classical field $\int_{0}^{t}v_s^C ds$ is represented by
the unitary Weyl operator $\hat W(v^C_{t]})$
(here $v^C_{t]}(s)=v_s^C$ for $0\leq s\leq t$, and 0 otherwise)
satisfying the quantum stochastic differential equation (QSDE):
\begin{eqnarray*}
    d\hat W(v_{t]}^C)&=&(v_{t]}^Cd\hat B_{o,t}^* -v_{t]}^{C*} d\hat B_{o,t}
          -\frac{1}{2}|v_{t]}^C|^2dt)\hat W(v_{t]}^C);\\
&&
    \hat W(v_{0]}^C)=I,
\end{eqnarray*}
with which we can write
$\hat u_{c,t}=\hat W(v_{t]}^C)^*\hat B_{o,t}\hat W(v_{t]}^C)$.
The coherent field $\hat u_{c,t}$ then interacts with the cavity
via mirror $M_2$ with coupling coefficient $\sqrt{\gamma}$, thus
the total cavity-fields interaction is described by the following QSDE:
\begin{eqnarray*}
& & \hspace*{-1em}
    d\hat U_t=\Big[-\frac{\kappa}{2}\hat a^* \hat adt
              +\sqrt{\kappa}(\hat a d\hat B_{in,t}^*
                            -\hat a^* d\hat B_{in,t})
\nonumber \\ & & \hspace*{4em}
              -\frac{\gamma}{2}\hat a^* \hat adt
              +\sqrt{\gamma}(\hat a d\hat u_{c,t}^*
                            -\hat a^* d\hat u_{c,t})
                \Big]\hat U_t.
\end{eqnarray*}
For a sufficiently small value of $\gamma$, the effect of the noise
$\hat B_{o,t}$ can be considered to be negligible and if also
$\gamma \ll \kappa$ then its contribution to the system noise will
be negligible compared to that of $\hat B_{in,t}$.
As a result, we find that the feedback term is included in the
interaction:
\begin{equation*}
    d\hat U_t=\Big[-\im\hat Fdt
               -\frac{\kappa}{2}\hat a^* \hat adt
              +\sqrt{\kappa}(\hat a d\hat B_{in,t}^*
                            -\hat a^* d\hat B_{in,t})
                \Big]\hat U_t.
\end{equation*}
The entire scheme is depicted in Fig.~\ref{fg_cav_diss}.
Note that a pumped $\chi^{(2)}$ nonlinear crystal can be placed
inside the cavity to implement these linear couplings together
with the quadratic Hamiltonian discussed in the preceding subsection.

%\begin{figure}[t]
%\centering
%\begin{picture}(200,100)
%\put(-20,5)
%{\includegraphics[width=3.4in]{cav_imp_cntl.png}}
%\put(-19,67){$\hat{B}_{in,t}$}
%\put(-19,21){$\hat{B}_{out,t}$}
%\put(19,45){$\hat{a}=\frac{\hat{q}+i\hat{p}}{\sqrt{2}}$}
%\put(62,63){$\hat{u}_{c,t}$}
%\put(142,55){$\tilde{f}$}
%\put(182,56){$1/\sqrt{\gamma}$}
%\put(212,65){$u_t$}
%\put(5,11){$\sqrt{\kappa}$}
%\put(52,11){$\sqrt{\gamma}$}
%\put(55,89){Control}
%\put(58,79){field}
%\put(84,55){MOD}
%\end{picture}
%\caption{
%\label{fg_cav_diss}
%Implementation of a dissipative coupling together with
%the linear controls.
%}
%\end{figure}

%%%%%%%%%%%%%%%%%%%%%%%%%%%%%%%%%%%%%%%%%%%%%%%%%%%%%%%%%%%%%%%%%%%%%

\subsection{Models with dispersive coupling
$\hat L=\sqrt{\kappa} \hat q$ and direct measurement feedback}

\begin{figure}[th]
\centering
\includegraphics[scale=0.65]{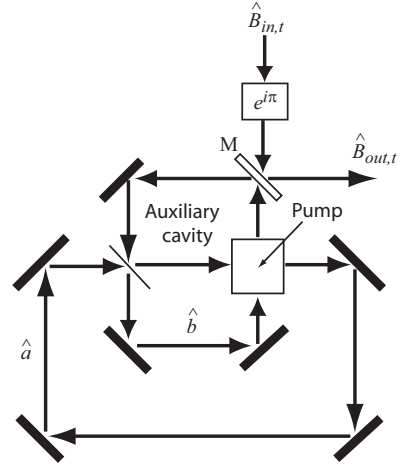}
\caption{Configuration of two cavities, a two mode squeezer
(depicted by the square with an arrow to indicate classical
pumping), and a beam splitter, to implement a dispersive coupling
of the cavity mode $\hat a$ to the field $\hat B_{in,t}$ when the
second cavity mode $\hat b$ is adiabatically eliminated. Note that
$\hat B_{in,t}$ is phase shifted by 180$^{\circ}$ before reaching
the mirror $M$.}
\label{fg_disp_coup}
\end{figure}

For realization of a dispersive coupling of the form $\hat
L=\sqrt{\kappa} \hat q$, consider the configuration shown in
Fig.~\ref{fg_disp_coup}. This configuration consists of a ring
cavity with mode $\hat a$, an auxiliary ring cavity with mode
$\hat b$, a $\chi^{(2)}$ nonlinear crystal in which the cavity
modes $\hat a$ and $\hat b$ interact with a classical pump beam,
and a beam splitter. The frequency of the auxiliary cavity is
matched to half the frequency of the pump beam. The  classically
pumped nonlinear crystal implements a two-mode squeezing
Hamiltonian given by $\hat H_{TMS}=\frac{\im}{2}(\epsilon
e^{-\im\omega_p t} \hat a^*\hat b^*- \epsilon^* e^{\im\omega_p t}\hat
a \hat b)$ (where $\epsilon$ is the effective intensity of the
classical pump and $\omega_p$ is the pump frequency), while the
beam splitter implements the Hamiltonian $\hat H_{BS}=\alpha \hat
a^*\hat b+\alpha^*\hat a \hat b^*$ for a complex parameter
$\alpha$. Suppose now that $\epsilon$ and $\alpha$ are chosen to
satisfy $\epsilon/2=\im\alpha=\Gamma$ for a real constant $\Gamma>0$
(in particular $\alpha=-\im\Gamma$). Then in a rotating frame at the
frequency $\omega_p/2$, the overall interaction Hamiltonian
between the modes $\hat a$ and $\hat b$ is thus given by
\[
   \hat H_{\hat a \hat b}
           =i\Gamma (\hat a^*\hat b^*-\hat a \hat b -
                 \hat a^* \hat b+\hat a \hat b^*).
\]
%

%\begin{figure}[t]
%\centering
%\begin{picture}(200,200)

%\put(20,5)
%{\includegraphics[width=2.2in]{cav_imp_disp.png}}

%\put(97,180){$\hat{B}_{in,t}$}
%\put(147,138){$\hat{B}_{out,t}$}
%\put(120,157){$e^{i\pi}$}
%\put(100,134){$M$}
%\put(25,45){$\hat{a}$}
%\put(94,58){$\hat{b}$}
%\put(77,110){Auxiliary}
%\put(80,96){Cavity}
%\put(138,111){Pump}

%\end{picture}
%\caption{
%\label{fg_disp_coup}
%Configuration of two cavities, a two mode squeezer
%(depicted by the square with an arrow to indicate classical
%pumping), and a beam splitter, to implement a dispersive coupling
%of the cavity mode $\hat a$ to the field $\hat B_{in,t}$ when the
%second cavity mode $\hat b$ is adiabatically eliminated. Note that
%$\hat B_{in,t}$ is phase shifted by 180$^{\circ}$ before reaching
%the mirror $M$.
%}
%\end{figure}

Assume that the coupling coefficient of the mirror $M$ is large
such that the mode $\hat b$ is heavily damped compared to mode
$\hat a$ and has much faster dynamics than $\hat a$. For
simplicity, in the following we will use the formalism of quantum
Langevin equations and a formal method to show that the
configuration shown implements the dispersive coupling $\hat L$ in
the reduced dynamics for mode $\hat a$ only, after mode $\hat b$
is adiabatically eliminated 
\footnote{As pointed out in \cite{GvH07}, in general one has 
to be careful when using such a formal method. 
However, in the particular case considered here
where the dynamics are linear it does turn out that the
formal method gives a consistent result in that the adiabatically
eliminated system is a bona fide quantum mechanical system.
}. 
See \cite{NJD08} for a more rigorous derivation using QSDEs and 
the mathematical theory for adiabatic elimination developed in 
\cite{BvHS07}.

Let $\hat B_{in,t}$ be an input field and $\hat B_{out,t}$ be an
output field coupled to $\hat b$ at the mirror $M$ as shown in
Fig.~\ref{fg_disp_coup} and suppose that $M$ has the coupling
coefficient $\gamma$. In particular, notice the 180$^{\circ}$
phase shift in front of $\hat B_{in,t}$ before it strikes the
mirror. Define $\hat \eta$ to be a quantum white noise formally
related to $\hat B_{in,t}$ as $\hat B_{in,t}=\int_{0}^t \hat
\eta(s)ds$ and $\hat \eta_{out}$ be the output noise at the mirror
$M$ formally related to $\hat B_{out,t}$ by $\hat
B_{out,t}=\int_{0}^t \hat \eta_{out}(s)ds$. The quantum Langevin
equations for the dynamics of $\hat a$, $\hat b$, and $\hat
\eta_{out}$ are \cite[Chapter 5]{Gardiner}
\begin{eqnarray}
  \dot{\hat a}
       &=&\im[\hat H_{\hat a\hat b},\hat a]
\nonumber\\
       &=&\Gamma (\hat b^*-\hat b),
\label{eq:app-a-dyn}\\
  \dot{\hat b}
       &=&\im[\hat H_{\hat a \hat b},\hat b]
              -\frac{\gamma}{2}\hat b+\sqrt{\gamma}\hat \eta
\nonumber\\
       &=&\Gamma (\hat a^*+\hat a)
              -\frac{\gamma}{2}\hat b+\sqrt{\gamma}\hat \eta,
\label{eq:app-b-dyn}\\
       \hat \eta_{out}&=&\sqrt{\gamma}\hat b-\hat \eta.
\label{eq:app-bout-dyn}
\end{eqnarray}
Setting $\dot{\hat b}=0$ and solving Eq. (\ref{eq:app-b-dyn}) for
$\hat b$ in terms of $\hat a^*$, $\hat a$ and $\hat \eta$ we
obtain
\begin{equation}
   \hat b = \frac{2}{\gamma}\left(\Gamma(\hat a^*+\hat a)
               +\sqrt{\gamma}\hat \eta\right).
\label{eq:app-b-elim}
\end{equation}
Substituting Eq. (\ref{eq:app-b-elim}) into Eqs.
(\ref{eq:app-a-dyn}) and (\ref{eq:app-bout-dyn}) we obtain that
the reduced dynamics for $\hat a$ only and $\hat \eta_{out}$ are
given by the following quantum Langevin equations:
\begin{eqnarray}
   \dot{\hat a}&=& \Gamma (\hat b^*-\hat b)
\nonumber\\
               &=& \frac{2\Gamma}{\sqrt{\gamma}}(-\hat \eta+\hat \eta^*),
\label{eq:app-a-fin}\\
   \hat \eta_{out}&=&\frac{2\Gamma}{\sqrt{\gamma}}(\hat a + \hat a^*) +
                         \hat \eta.
\label{eq:app-y-fin}
\end{eqnarray}
The pair of equations (\ref{eq:app-a-fin}) and
(\ref{eq:app-y-fin}) shows that the reduced system after adiabatic
elimination of the mode $\hat b$ is a single degree of freedom
quantum harmonic oscillator with mode $\hat a$ coupled to the
field $\hat B_{in,t}$ by the linear coupling operator $\hat
L=2\sqrt{2}\Gamma\hat q/\sqrt{\gamma}$, where $\hat q=(\hat a+\hat
a^*)/\sqrt{2}$, producing the output field $\hat
B_{out,t}=\int_{0}^t \hat \eta_{out}(s)ds$. By suitably choosing
$\Gamma$ and $\gamma$ such that
$2\sqrt{2}\Gamma=\sqrt{\kappa\gamma}$ we see that with this scheme
it is possible to implement any dispersive coupling of the form
$\hat L=\sqrt{\kappa}\hat q$. Moreover, by placing a pumped
nonlinear crystal inside the cavity (pumped with the same
frequency $\omega_p$) and adding a partially transmitting mirror
in the ring cavity of $\hat a$ that couples it to a control field,
one can easily combine this dispersive coupling together with the
quadratic Hamiltonian in Subsection 1 of this Appendix as well as
the control shown in Fig.~\ref{fg_cav_diss}.

%%%%%%%%%%%%%%%%%%%%%%%%%%%%%%%%%%%%%%%%%%%%%%%%%%%%%%%%%%%%%%%%%%%%%%%%%
%%%%%%%%%%%%%%%%%%%%%%%%%%%%%% References %%%%%%%%%%%%%%%%%%%%%%%%%%%%%%%
%%%%%%%%%%%%%%%%%%%%%%%%%%%%%%%%%%%%%%%%%%%%%%%%%%%%%%%%%%%%%%%%%%%%%%%%%


\begin{thebibliography}{99}



\bibitem{nielsen}
M. A. Nielsen and I. L. Chuang,
{\it Quantum Computation and Quantum Information},
(Cambridge University Press, Cambridge, 2000).

\bibitem{Cirac}
J. I. Cirac, P. Zoller, J. H. Kimble, and H. Mabuchi,
%Quantum state transfer and entanglement distribution among distant
%nodes in a quantum network,
Phys. Rev. Lett. 78, 3221 (1997).

\bibitem{vanEnk}
S. J. van Enk, J. I. Cirac, and P. Zoller,
%Ideal quantum-communication over noisy channels: a quantum
%optical implementation,
Phys. Rev. Lett. 78, 4293 (1997).

\bibitem{Parkins1}
A. S. Parkins and H. J. Kimble,
%Quantum state transfer between motion and light,
J. Opt. B: Quantum Semiclass. Opt. 1, 496 (1999).

\bibitem{Parkins2}
A. S. Parkins and H. J. Kimble,
%Position-momentum Einstein-Podolsky-Rosen state of distantly
%separated trapped atoms,
Phys. Rev. A 61, 052104 (2000).

%\bibitem{shor}
%P. Shor,
%Scheme for reducing decoherence in quantum computer memory,
%Phys. Rev. A 52, 2493 (1995).

%\bibitem{steane}
%A. M. Steane,
%Error-correcting codes in quantum theory,
%Phys. Rev. Lett. 77, 793 (1996).

%\bibitem{knill}
%E. Knill and R. Laflamme,
%Theory of quantum error-correcting codes,
%Phys. Rev. A 55, 900 (1997).

\bibitem{Bennett1}
C. H. Bennett, G. Brassard, S. Popescu, B. Schumacher, J. A. Smolin,
and W. K. Wootters,
%Purification of Noisy Entanglement and Faithful Teleportation
%via Noisy Channels,
Phys. Rev. Lett. 76, 722 (1996).

\bibitem{Bennett2}
C. H. Bennett, D. P. DiVincenzo, J. A. Smolin, and W. K. Wootters,
%Mixed-state entanglement and quantum error correction,
Phys. Rev. A 54, 3824 (1996).

\bibitem{Eberly1}
T. Yu and J. H. Eberly,
%Finite-time disentanglement via spontaneous emission,
Phys. Rev. Lett. 93, 140404 (2004).

\bibitem{Eberly2}
J. H. Eberly and T. Yu,
%The end of an entanglement,
Science, 316, 27 (2007).

\bibitem{Doherty1}
A. C. Doherty and K. Jacobs,
%Feedback control of quantum systems using continuous state estimation,
Phys. Rev. A 60, 2700 (1999).

\bibitem{Thomsen}
L. Thomsen, S. Mancini, and H. M. Wiseman,
%Continuous quantum nondemolition feedback and unconditional atomic spin
%squeezing,
J. Phys. B, 35, 4937 (2002).

\bibitem{Ahn}
C. Ahn, H. M. Wiseman, and G. J. Milburn,
%Quantum error correction for continuously detected errors,
Phys. Rev. A 67, 052310 (2003).

\bibitem{Bouten}
L. M. Bouten, R. van Handel, and M. R. James,
%A discrete invitation to quantum filtering and feedback control,
to appear in SIAM Review, arXiv: math/0606118 (2006).

\bibitem{Wang}
J. Wang, H. M. Wiseman, and G. J. Milburn,
%Dynamical creation of entanglement by homodyne-mediated feedback
Phys. Rev. A 71, 042309 (2005).

\bibitem{Carvalho}
A. R. R. Carvalho and J. J. Hope,
%Stabilizing entanglement by quantum-jump-based feedback,
Phys. Rev. A 76, 010301(R) (2007).

\bibitem{Mirrahimi}
M. Mirrahimi and R. van Handel,
%Stabilizing feedback controls for quantum systems,
{\it SIAM J. Control Optim.} 46, 445-467 (2007).

\bibitem{Yamamoto}
N. Yamamoto, K. Tsumura, and S. Hara,
%Feedback control of quantum entanglement in a two-spin system,
{\it Automatica} 43, 981-992 (2007).

\bibitem{Mancini}
S. Mancini and H. M. Wiseman,
%Optimal control of entanglement via quantum feedback,
Phys. Rev. A 75, 012330 (2007).

\bibitem{Wiseman1}
H. M. Wiseman and G. J. Milburn,
%Quantum theory of optical feedback via homodyne detection,
Phys. Rev. Lett. 70, 548 (1993).

\bibitem{Wiseman2}
H. M. Wiseman,
%Quantum theory of continuous feedback,
Phys. Rev. A 49, 2133 (1994).

\bibitem{Yanagisawa}
M. Yanagisawa,
%Quantum feedback control for deterministic entanglement photon generation,
Phys. Rev. Lett. 97, 190201 (2006).

\bibitem{Warszawski1}
P. Warszawski, H. M. Wiseman, and H. Mabuchi,
%Quantum trajectories for realistic detection,
Phys. Rev. A 65, 023802 (2002).

\bibitem{Warszawski2}
P. Warszawski and H. M. Wiseman,
%Quantum trajectories for realistic detection,
J. Opt. B: Quantum Semiclass. Opt. 5, 1 (2003).

\bibitem{Warszawski3}
P. Warszawski and H. M. Wiseman,
%Quantum trajectories for realistic detection,
J. Opt. B: Quantum Semiclass. Opt. 5, 15 (2003).

\bibitem{Braunstein}
S. L. Braunstein and P. van Loock,
%Quantum information with continuous variables,
Rev. Mod. Phys. 77, 513 (2005).

\bibitem{Carmichael1}
H. J. Carmichael,
{\it An open system approach to quantum optics},
Springer, Berlin (1993).

\bibitem{Carmichael2}
H. J. Carmichael,
%Quantum trajectory theory for cascaded open systems,
Phys. Rev. Lett. 70, 2273 (1993).

\bibitem{Gardiner}
C. W. Gardiner and P. Zoller,
{\it Quantum Noise},
Springer, Berlin (2000).

\bibitem{Gough}
J. Gough and M. R. James,
%The series product and its application to quantum feedforward and
%feedback networks,
arXiv: 0707.0048 (2007).

\bibitem{Hudson}
R. L. Hudson and K. R. Parthasarathy,
%Quantum Ito's formula and stochastic evolution,
Commun. Math. Phys. 93, 301 (1984).

\bibitem{Edwards}
S. C. Edwards and V. P. Belavkin,
arXiv: quant-ph/0506018 (2005).

\bibitem{NJD08}
H. I. Nurdin, M. R. James, and A. C. Doherty,
%Network synthesis of linear dynamical quantum stochastic
%systems,
arXiv:0806.4448 (2008).

\bibitem{JNP06}
M. R. James, H. I. Nurdin, and I. R. Petersen,
to appear in IEEE Trans. Automat. Contr.,
arXiv:quant-ph/0703150 (2007).

\bibitem{Duan}
L. M. Duan, G. Giedke, J. I. Cirac, and P. Zoller,
%Inseparability criterion for continuous variable systems,
Phys. Rev. Lett. 84, 2722 (2000).

\bibitem{Simon}
R. Simon,
%Peres-Horodecki separability criterion for continuous variable systems,
Phys. Rev. Lett. 84, 2726 (2000).

\bibitem{Vidal}
G. Vidal and R. F. Werner,
%Computable measure of entanglement,
Phys. Rev. A 65, 032314 (2002).

\bibitem{Zhou}
K. Zhou, J. Doyle, and K. Glover,
{\it Robust and Optimal Control},
Prentice Hall, NJ (1996).

\bibitem{Ficek}
Z. Ficek and R. Tanas,
%Delayed sudden-birth of entanglement,
Phys. Rev. A 77, 054301 (2008).

\bibitem{Paris}
M. G. A. Paris, F. Illuminati, A. Serafini, and S. DeSiena,
%Purity of Gaussian states: Measurement scheme and time evolution
%in noisy channel,
Phys. Rev. A 68, 012314 (2003).

\bibitem{Milburn}
G. J. Milburn,
%Classical and quantum conditional statistical dynamics,
Quantum Semiclass. Opt. 8, 269 (1996).

\bibitem{Doherty2}
A. C. Doherty, S. M. Tan, A. S. Parkins, and D. F. Walls,
%State determination in continuous measurement,
Phys. Rev. A 60, 2380 (1999).

\bibitem{Wiseman4}
H. M. Wiseman and G. J. Milburn,
%Quantum theory of field-quadrature measurements,
Phys. Rev. A 47, 642 (1993).

\bibitem{Wiseman3}
H. M. Wiseman and A. C. Doherty,
%Optimal unravellings for feedback control in linear quantum systems,
Phys. Rev. Lett. 94, 070405 (2005).

\bibitem{DHJ06}
C. {D'Helon} and M. R. James, 
Phys. Rev. A 73, 053803 (2006).

\bibitem{BvHS07}
L. Bouten, R. van Handel, and A. Silberfarb,
%Approximation and limit theorems for quantum stochastic models
%with unbounded coefficients,
Journal of Functional Analysis 254, 3123 (2008).

\bibitem{GvH07}
J. Gough and R. van Handel,
J. Stat. Phys. 127, 575 (2007).


\end{thebibliography}
\end{document}